%% file: main.tex
\documentclass[acmsmall]{acmart}

\usepackage{soul}
\usepackage{xcolor}

\definecolor{ao(english)}{rgb}{0.0, 0, 1}

\newcommand{\add}[1]{\textcolor{black}{#1}}

\newcommand{\remove}[1]{}

\newcommand{\addminor}[1]{\textcolor{black}{#1}}

\newcommand{\removeminor}[1]{}

\AtBeginDocument{%
  \providecommand\BibTeX{{%
    \normalfont B\kern-0.5em{\scshape i\kern-0.25em b}\kern-0.8em\TeX}}}
\AtBeginEnvironment{quote}{\itshape}

\usepackage{siunitx}
\usepackage{booktabs}
\usepackage{multirow}
\usepackage[normalem]{ulem}
\useunder{\uline}{\ul}{}
\begin{document}

%%
%% The "title" command has an optional parameter,
%% allowing the author to define a "short title" to be used in page headers.
\title[AppealMod: Inducing Friction to Reduce Moderator Workload of Handling User Appeals]{\addminor{AppealMod: Inducing Friction to Reduce Moderator Workload of Handling User Appeals} \removeminor{AppealMod: Shifting Effort from Moderators to Users Making Appeals} }

%%
%% The "author" command and its associated commands are used to define
%% the authors and their affiliations.
%% Of note is the shared affiliation of the first two authors, and the
%% "authornote" and "authornotemark" commands
%% used to denote shared contribution to the research.
\author{Shubham Atreja}
\email{satreja@umich.edu}
\affiliation{
  \institution{University of Michigan School of Information}
%   \streetaddress{P.O. Box 1212}
  \city{Ann Arbor}
  \state{Michigan}
  \country{USA}
%   \postcode{43017-6221}
}

\author{Jane Im}
\email{imjane@umich.edu}
\affiliation{
  \institution{University of Michigan School of Information and the Division of Computer Science \& Engineering}
%   \streetaddress{P.O. Box 1212}
  \city{Ann Arbor}
  \state{Michigan}
  \country{USA}
%   \postcode{43017-6221}
}

\author{Paul Resnick}
\email{presnick@umich.edu}
\affiliation{
  \institution{University of Michigan School of Information}
%   \streetaddress{P.O. Box 1212}
  \city{Ann Arbor}
  \state{Michigan}
  \country{USA}
%   \postcode{43017-6221}
}

\author{Libby Hemphill}
\email{libbyh@umich.edu}
\affiliation{
  \institution{University of Michigan School of Information and ICPSR}
%   \streetaddress{P.O. Box 1212}
  \city{Ann Arbor}
  \state{Michigan}
  \country{USA}
%   \postcode{43017-6221}
}

%%
%% By default, the full list of authors will be used in the page
%% headers. Often, this list is too long, and will overlap
%% other information printed in the page headers. This command allows
%% the author to define a more concise list
%% of authors' names for this purpose.
\renewcommand{\shortauthors}{Atreja et al.}

\setcopyright{acmlicensed}
\acmJournal{PACMHCI}
\acmYear{2024} \acmVolume{8} \acmNumber{CSCW1} \acmArticle{19} \acmMonth{4} \acmPrice{15.00}\acmDOI{10.1145/3637296}

%%
%% The abstract is a short summary of the work to be presented in the
%% article.
\input{content/abstract.tex}



%%
%% Keywords. The author(s) should pick words that accurately describe
%% the work being presented. Separate the keywords with commas.

\received{January 2023}
% \received[revised]{October 2023}
\received[accepted]{December 2023}

% \received{20 February 2007}
% \received[revised]{12 March 2009}
% \received[accepted]{5 June 2009}

%%
%% This command processes the author and affiliation and title
%% information and builds the first part of the formatted document.
\maketitle
% \forlater{Looking for a new title; suggestions welcome} \jane{I think adding a keyword about users' effort is important... Something like: 
% "AppealMod: A System for Assisting Moderators Address User Appeals By Creating Effort-based Friction"} 

\noindent \emph{\textbf{CONTENT WARNING:} This paper contains offensive language, including misogynistic slurs, that readers may find disturbing.}

\input{content/intro}

\input{content/background}

\input{content/design}
\input{content/method}
\input{content/results}
\input{content/limitations}
\input{content/discussion}

\bibliographystyle{ACM-Reference-Format}
\bibliography{sample-base}

\input{content/z_appendix}

\end{document}

%% file: content/abstract.tex
\begin{abstract}
    As content moderation becomes a central aspect of all social media platforms and online communities, interest has grown in how to make moderation decisions contestable. On social media platforms where individual communities moderate their own activities, the responsibility to address user appeals falls on volunteers from within the community. While there is a growing body of work devoted to understanding and supporting the volunteer moderators' workload, little is known about their practice of handling user appeals. Through a collaborative and iterative design process with Reddit moderators, we found that moderators spend considerable effort in investigating user ban appeals and desired to directly engage with users and retain their agency over each decision. To fulfill their needs, we designed and built AppealMod, a system that induces \emph{friction} in the appeals process by asking users to provide additional information before their appeals are reviewed by human moderators. In addition to giving moderators more information, we expected the friction in the appeal process would lead to a selection effect among users, with many insincere and toxic appeals being abandoned before getting any attention from human moderators. To evaluate our system, we conducted a \add{randomized} field experiment in a Reddit community of over 29 million users that lasted for four months. As a result of the selection effect, moderators viewed only 30\% of initial appeals and less than 10\% of the toxically worded appeals; yet they granted roughly the same number of appeals \add{when compared with the control group}. Overall, our system is effective at reducing moderator workload and minimizing their exposure to toxic content while honoring their preference for direct engagement and agency in appeals. 
    
    % Users with legitimate appeals did provide the extra information; the total number of appeals granted in the treatment and control conditions was similar.

\end{abstract}

%% file: content/intro.tex
\section{Introduction}

% set the stage around contestability of content moderation decisions

As content moderation becomes a central aspect of all social media platforms and online communities, interest has grown in how to make moderation decisions contestable \cite{vaccaroContestabilityContentModeration2021a,vaccaroEndDayFacebook2020}. Particularly as certain decisions (e.g., permanently banning users) can have long-term consequences, moderation systems need to ensure they allow users to appeal individual decisions in case of error or injustice. 
%can lead to significant negative outcomes (e.g., suppressing minority voices, limiting access to work opportunities), 
% Facebook \cite{nprFacebookUpdates}, Instagram \cite{engadgetInstagramWill}, and Twitter \cite{techcrunchTwitterLets} have all recently introduced or updated their processes for appealing moderation decisions. 
For platforms and communities, appeals provide a mechanism to evaluate individual decisions. For users, they offer opportunities to understand the moderation policies and modify their behavior  \cite{myerswestCensoredSuspendedShadowbanned2018b}. 

% dive into volunteer moderators and their workload. 

On social media platforms that empower individual communities to manage and moderate their own activities, the responsibility to address user appeals falls on volunteers within the community \cite{gillespieCustodiansInternetPlatforms2018,schopke-gonzalezWhyVolunteerContent2022}. These communities include groups on Facebook, subreddits on Reddit, and channels on Twitch and YouTube. Within these communities, volunteer moderators perform all moderation tasks from formulating and enforcing rules to explaining their decisions and considering appeals \cite{wohnVolunteerModeratorsTwitch2019,seeringModeratorEngagementCommunity2019b}. Consequently, a growing body of content moderation research \cite{jhaverHumanMachineCollaborationContent2019b,chandrasekharanCrossmodCrossCommunityLearningbased2019b,caiCategorizingLiveStreaming2019a,liAllThatHappening2022,seering2023moderates, zhangPolicyKitBuildingGovernance2020b} focuses on understanding and supporting volunteer moderators' work. For instance, some research aims to help moderators identify and remove problematic content \cite{chandrasekharanCrossmodCrossCommunityLearningbased2019b,jhaverDesigningWordFilter2022}. 
% For instance, researchers have developed an assortment of tools -- word-filters \cite{jhaverDesigningWordFilter2022}, machine learning classifiers \cite{chandrasekharanCrossmodCrossCommunityLearningbased2019b}, and forecasting models -- to help moderators identify and remove problematic content, as well as to help them formulate moderation policies \cite{zhangPolicyKitBuildingGovernance2020b}.
However, other aspects of volunteer moderators' work, particularly how they handle user appeals, have received little attention.  

%  start listing down more specific challenges. 

To understand moderators' handling of user appeals, we turned to prior work on social media users' contestability needs \cite{vaccaroContestabilityContentModeration2021a,vaccaroEndDayFacebook2020}. For instance, \citet{vaccaroEndDayFacebook2020} found that when users contest moderation decisions, they almost always request a human review and prefer to directly engage with the moderators. However, directly communicating with users on every appeal can amount to a significant workload for volunteer moderators who are already overwhelmed with their work, and suffer occupational stress and burnout \cite{schopke-gonzalezWhyVolunteerContent2022, dosonoModerationPracticesEmotional2019}. Directly engaging with users also puts moderators at an increased risk of being exposed to toxic content from community members who are angry or upset about their decision \cite{gilbertRunWorldLargest2020a,wohnVolunteerModeratorsTwitch2019}. Ideally, a successful appeal process would honor both users' and moderators' needs.

In this paper, we investigated Reddit volunteer moderators' current practices of handling appeals from users who are banned from their community, and worked with them to design a system to help them process appeals fairly and efficiently. On Reddit, volunteer moderators cannot control who joins their community, but they have the power to ban members. As a result, moderators actively use banning as a strategy to keep problematic users away from their community \cite{seeringModeratorEngagementCommunity2019b}. Furthermore, given the long-term consequences of a permanent ban, Reddit requires that moderators must allow users to appeal bans.  

Through a collaborative and iterative design process with Reddit moderators, we identified their strategies and needs for addressing appeals from banned users. Our interviews and design sessions with mod\add{erator}s revealed an effort asymmetry between moderators and users. While users could submit an appeal with minimal effort using Reddit's current system, moderators described spending considerable time and effort investigating and reviewing these appeals. They assessed whether a banned user had reflected on their behavior and tested the user's awareness of community norms. Moderators also desired to directly engage with banned users, despite the increased risk of receiving toxic messages, and to retain their agency over final decisions. 

% \jane{[Quick note that I think it could be more convincing to introduce your idea of friction-based moderation first, and then say that AppealMod is an instantiation of it.]} 

To fulfill these needs, we created AppealMod, a system that helps Reddit moderators to process appeals from banned users. AppealMod shifts the onus to banned users to complete their appeal by answering additional questions about their past behavior and their understanding of the community norms. To reduce the burden on moderators, appeals are hidden from their view until after users complete the AppealMod process. Completing the process restores the direct engagement channel between banned users and moderators, who then make the final decision. By asking users to put more effort into their appeal initially, AppealMod effectively induces \emph{friction} in the appealing process. We expected this friction to lead to a selection effect among users as some users would be discouraged by the AppealMod process, ideally exactly those users who had insincere or toxically worded appeals. We also hoped that the extra information provided by those appellants who completed the process would save effort for the moderators in assessing the appeals. 

We conducted a field experiment to evaluate AppealMod in r/pics, a subreddit of over 29 million users.\footnote{As of January 2023, the community had 29.7 million users.} The experiment lasted for four months. Appealing users were randomly assigned to a control or a treatment condition. Under control, users followed Reddit's existing process to submit their appeal. Users under treatment were subject to the AppealMod process for submitting their appeal. Results from the experiment show that roughly 70\% of appealing users under treatment were discouraged by the AppealMod process and abandoned their appeal. More specifically, we found that users making insincere appeals or using toxic language were more likely to abandon the process. Therefore, while moderators reviewed only 30\% of the appeals under treatment, they still granted roughly the same number of appeals under control and treatment. AppealMod also offered moderators some protection from toxic content. 91.3\% of the toxic appeals subject to the AppealMod process were abandoned and remained hidden from the moderators. We conclude the paper by discussing implications for inducing friction in moderation processes to reduce the workload of volunteer moderators and potential next steps for improving the design of AppealMod. 

% Our design demonstrates the potential of pricing-based approaches for reducing  We believe that pricing-based approaches will be most effective when they are perceived as useful by deserving users. Therefore, it is important to inform the design by moderators' existing processes to design the process in close collaboration with moderators, and to evaluate their design a in real setting to account pricing-based approaches in content moderation and the design of contestability systems more broadly. 

%% file: content/background.tex
\section{Background and Related Work}
% notes from discussion: effort asymmetry and pricing; moderators' and users' needs in contesting decisions
We first review prior research on volunteer moderators' experiences and needs, as well as existing work on designing tools for moderators. We then review research on contestability of content moderation decisions. Lasty, we discuss costs and friction-based techniques to reduce volunteer moderators' workload.

\subsection{Volunteer Moderator Workload and Experiences}
% ## importance 
\citet{schopke-gonzalezWhyVolunteerContent2022} define volunteer content moderators as ``individuals who uphold their own online community's standards''. Many social media platforms empower members of individual communities to manage and moderate their own activities \cite{jiang2022trade}. This includes groups on Facebook, subreddits on Reddit, and streaming channels on Twitch and YouTube. Matias \cite{matiasCivicLaborVolunteer2019} has described the work of these volunteer moderators as a form of ``civic labor''. They are often members of the community who care about its growth and development \cite{seering2022metaphors}. These volunteers formulate the rules and standards of their community and also enforce them, which involves reviewing and removing potentially problematic content and users, explaining their decisions, and considering appeals on their decisions \cite{gillespieCustodiansInternetPlatforms2018, seeringModeratorEngagementCommunity2019b, wohnVolunteerModeratorsTwitch2019, liAllThatHappening2022}. As a result, volunteer moderators often find themselves overwhelmed with the varied demands of their work \cite{schopke-gonzalezWhyVolunteerContent2022, dosonoModerationPracticesEmotional2019}. 

In addition to being labor-intensive, moderating online communities can also take an emotional toll on the moderators \cite{roberts2019behind,steiger2021psychological,schopke-gonzalezWhyVolunteerContent2022}. When reviewing content, moderators have to look at some of the most toxic content the Internet has to offer, leading to significant psychological distress \cite{gilbertRunWorldLargest2020a, wohnVolunteerModeratorsTwitch2019}. Therefore, volunteer moderation has also been classified as a form of ``emotional labor'' \cite{dosonoModerationPracticesEmotional2019}. Furthermore, as volunteer moderators make decisions about what is permitted, they face harassment and abuse from community members who are angry or upset about a moderation decision \cite{loWhenAllYou2018, wohnVolunteerModeratorsTwitch2019}. 

While platforms are responsible for providing tools for volunteers to be able to perform moderation adequately, these tools often fall far short of addressing the moderators’ needs \cite{jhaverDesigningWordFilter2022, liAllThatHappening2022}. For example, the prolonged neglect of moderation software on Reddit was so frustrating that it pushed moderators to join together in protest against the platform \cite{matiasGoingDarkSocial2016}. In light of this, any tool that helps volunteer moderators manage their workload or offers protection from toxic content will likely both benefit  moderators and help the community grow. Our work builds on the existing literature to design and deploy a system that can address volunteer moderators' needs.

\subsection{Designing Tools for Volunteer Moderators}
\label{sec:background_tools}
% ## existing tools for moderators and the gap around contestability 

Given the inadequacy of platform-provided moderation tools, volunteer moderators increasingly rely on third-party tools, which include those developed by researchers. For instance, Automod, the most popular tool on Reddit, was first developed by a volunteer moderator and later adopted by Reddit \cite{jhaverHumanMachineCollaborationContent2019b}. Automod can be independently programmed by moderators to automatically detect (and act against) content that violates the rules of their community \cite{wright2022automated}. Relying on external tools is not unique to volunteer moderators on Reddit. \citet{caiCategorizingLiveStreaming2019a} found that streamers on Twitch also relied on third-party applications to accomplish their moderation tasks. The most common use of these tools included identifying potentially problematic content and taking action against problematic users, e.g., banning them \cite{caiCategorizingLiveStreaming2019a}.

Consequently, there is a growing thread of content moderation research focusing on the study and design of tools for volunteer moderators. This body of work has mainly resulted in an assortment of techniques and tools for detecting and acting against problematic content and users (e.g., word-filters \cite{jhaverDesigningWordFilter2022}, machine learning classifiers \cite{chandrasekharanCrossmodCrossCommunityLearningbased2019b}, forecasting tools \cite{liuForecastingPresenceIntensity2018}). In contrast, other aspects of volunteer moderators' work have received little attention. One notable exception is PolicyKit, developed by \citet{zhangPolicyKitBuildingGovernance2020b} to support moderators' task of managing and formulating their community rules. However, there has not yet been research on designing tools for supporting moderators' handling of user appeals, an essential task that volunteer moderators perform \cite{seeringModeratorEngagementCommunity2019b}.
In this work, we focus on moderators' handling of appeals from banned users and present a system aimed at supporting this workload. Banning is an important strategy that moderators use to remove problematic users from their community \cite{seeringModeratorEngagementCommunity2019b}, and given the long-term consequences of a permanent ban, 
it is essential that moderation systems allow users to appeal against their ban. 

% In this work, we focus on an unexplored but essential task that volunteer moderators perform -- addressing user appeals against their moderation decision -- and present a system aimed at reducing this workload. 
% \forlater{add more emphasis on why focusing on ban appeals is an important line of research}

% LH trying ot rewrite this paragraph

In order to be successful, tools for volunteer moderators must attend to how they are situated relative to moderators' existing work, processes, and control. Automating one aspect of the moderators' work may create manual work in other aspects of their role. For instance, \citet{jhaverHumanMachineCollaborationContent2019b} found that moderators' use of automated bots created new challenges of training and coordination among moderators and also added the new task of maintaining these bots \cite{jhaverHumanMachineCollaborationContent2019b}. Automation is also in tension with agency \cite{jiangTradeoffcenteredFrameworkContent2022,heerAgencyAutomationDesigning2019a,caiHumanCenteredToolsCoping2019a}. Moderators have resisted replacing rule-based tools with machine learning approaches \cite{jhaverDesigningWordFilter2022}, in part because they want to maintain agency over a system's decisions and impacts. They also find the ML systems difficult to understand and control, even when they outperform rule-based systems \cite{chandrasekharanCrossmodCrossCommunityLearningbased2019b}. 

% As \cite{caiHumanCenteredToolsCoping2019a, vealeFairnessAccountabilityDesign2018} explain, the tradeoff between agency and automation \cite{seering2020reconsidering} TKTK. 
%  sense of agency
%  NOT SURE IF WE NEED TO GET INTO THESE DETAILS:
% The moderators need to maintain agency is evident from their resistance to replacing rule-based tools with machine leaning based approaches\cite{word filters}. While machine learning approaches can often outperform rule-based approaches \cite{crossmod}, moderators find them difficult to understand and control.

% This has led some researchers to formulate an agency vs automation tradeoff \cite{24 from google paper, public stakeholders paper}. Yet, others have argued for rethinking the design of automation so that agency and automation are mutually beneficial \cite{heer, google paper}. 

% A common finding from research on tools for volunteer moderators is that these automation tools are ``deeply embedded into the social processes of moderation''~\cite{jhaverHumanMachineCollaborationContent2019b, seeringModeratorEngagementCommunity2019b}.  
% Furthermore, as members of the community, volunteer moderators are deeply invested in having control over their moderation operations \cite{jhaverDesigningWordFilter2022} and maintaining agency over decisions \cite{jiangTradeoffcenteredFrameworkContent2022}.  
% Such findings show the importance of the tools being well-integrated into volunteer moderation procedures.  

In our research, we adopted a collaborative and iterative design approach to create a tool informed by moderators’ existing practices and needs so that it can seamlessly integrate into their workflows.
We first built connections with volunteer moderators from different communities and invited them to participate in a research collaboration. We conducted a series of synchronous and asynchronous sessions to uncover their needs for addressing user ban appeals, including their need to maintain agency over each individual decision. We then designed a system to address their needs and conducted a field experiment to evaluate its effectiveness.

\subsection{Designing for Contestability of Moderation Decisions}

As we noted earlier, there is no prior work on understanding volunteer moderators' workload of handling user appeals. Designing for contestability from the users' point of view, however, is an important area of research \cite{vaccaroContestabilityContentModeration2021a,vaccaroEndDayFacebook2020}. Moderation systems can frequently make incorrect decisions \cite{holpuch2015facebook,propublicaFacebooksSecret}, and it is crucial for users to be able to appeal individual decisions as incorrect moderation decisions can cause significant negative outcomes for users \cite{outDangerousTrend} and social media platforms \cite{levy2014facebook} alike. Facebook \cite{nprFacebookUpdates}, Instagram \cite{engadgetInstagramWill}, and Twitter \cite{techcrunchTwitterLets} all recently introduced or updated their processes for appealing content moderation decisions. However, these systems are subject to frequent criticism from the users \cite{myerswestCensoredSuspendedShadowbanned2018b,vaccaroEndDayFacebook2020}. For instance, most appeal systems do not provide a space for users to explain their behavior \cite{vaccaroEndDayFacebook2020}. Some users describe the appeal process as ``speaking into a void'' \cite{myerswestCensoredSuspendedShadowbanned2018b} for its lack of direct human interaction. 

\add{Recently, researchers have started to explore alternate designs of appeal process.} \citet{vaccaroEndDayFacebook2020}\remove{, who explored alternate designs of appeal process,} found that when users have the option to add an explanation to their appeal, they actively use that space to defend their behavior, share additional context, or even raise questions about moderation policies. Researchers have also argued for ``scaffolding the appeal'' \cite{vaccaroEndDayFacebook2020}, i.e., the appeal process should make clear what information moderators would consider, for instance by providing a structured form, because many users do not know what to say in their appeals or how to persuade the moderators \cite{vaccaroEndDayFacebook2020}. Furthermore, by laying out factors that are considered in the decision-making, scaffolding can bring transparency to the process and reduce perceived inconsistencies in moderation decisions \cite{vaccaroContestabilityContentModeration2021a}. \add{At the same time, \citet{vaccaroEndDayFacebook2020}'s study did not find any differences in the subjects' fairness perceptions toward different appeal processes. However, it is worth noting that in their online experiment, participants were given hypothetical scenarios in which the participants’ written appeals could not change the final outcome---which remained undesirable for the user. Therefore, it is still worth evaluating alternative appeal processes in a real-world setting, especially as prior research has shown that moderation outcomes can impact people's satisfaction with the process \cite{pan2022comparing}.}
% -----original version------
% \citet{vaccaroEndDayFacebook2020}, who explored alternate designs of appeal process, found that when users have the option to add an explanation to their appeal, they actively use that space to defend their behavior, share additional context, or even raise questions about moderation policies. Researchers have also argued for ``scaffolding the appeal'' \cite{vaccaroEndDayFacebook2020}, i.e., the appeal process should make clear what information moderators would consider, for instance by providing a structured form, because many users do not know what to say in their appeals or how to persuade the moderators \cite{vaccaroEndDayFacebook2020}. Furthermore, by laying out factors that are considered in the decision-making, scaffolding can bring transparency to the process and reduce perceived inconsistencies in moderation decisions \cite{vaccaroContestabilityContentModeration2021a}. 
% \add{In \citet{vaccaroEndDayFacebook2020}'s lab-based evaluation of different appeal processes, the appeals written by users did not change their final outcome, which remained undesirable for the user. While their study did not find any differences in the subjects' fairness perceptions toward different appeal processes, it remains unclear how that will change in a real world setting, especially as some appeal processes can lead to more favorable outcomes \cite{pan2022comparing}.}

Prior work also notes that social media users considered communication, especially direct communication with human moderators, as one of the top three avenues for improving their abilities to contest content moderation decisions \cite{vaccaroContestabilityContentModeration2021a}. When writing their appeals, many users explicitly requested a human review \cite{vaccaroEndDayFacebook2020}. \add{The requirement of a human review is becoming increasingly important as more decisions are taken via automated systems. Both scholarly and legal frameworks have underlined the importance of human review in their conceptualization of contestability \cite{lyons2021conceptualising, almada2019human}. For instance, 
\citet{sarra2020put} advocates for contestability in the form of dialectical exchanges between users and decision systems that allow users to ask questions and collect any additional information needed to contest their decisions. 
However, challenges still remain in terms of the scaling of human review while also tackling bias and decision fatigue \cite{lyons2021conceptualising}.
}

In sum, prior research has underlined the importance of allowing users to contest moderation decisions and provided multiple design recommendations for supporting users' contestability needs. While we designed AppealMod to fulfill moderators' needs while addressing user appeal, 
our system nevertheless implements many of the design recommendations that can support users' contestability needs as well. In particular, AppealMod provides scaffolding to users by sharing a webform with questions that moderators would consider in their decision making. Furthermore, all users who complete the AppealMod process get an opportunity to interact with the moderators and have their appeals \add{individually} reviewed \add{and responded to} by a human.

\subsection{Cost and Friction for Reducing Problematic User Behavior in Content Moderation}

Part of what makes handling user appeals challenging for moderators is the volume of appeals submitted. For instance, in 2021, moderators permanently banned over 4.9 million users from various communities on Reddit \cite{redditincTransparencyReport}. Even a small proportion of users (say 5\%) appealing their decision will result in roughly 250,000 appeals for the moderators. 
% In 2021, Reddit admins received 79,009 appeals, among which  80.35\% were denied \cite{redditincTransparencyReport}. 
% For instance, in 2021, moderators permanently banned over 4.9 million users from various communities on Reddit \cite{redditincTransparencyReport}. 
One way to reduce the volume of a particular behavior is to make it costly via direct monetary payments \cite{enders2008long,oestreicher2013content}. For instance, an online community could require an entry fee to join the community or a fee to submit an appeal of a decision. This is different from the fees charged for an online service, such as streaming platforms \cite{swatman2006changing}, in that the appeal fee is intended to make users recognize the cost of their actions, and thereby reduce problematic behaviors~\cite{grimmelmannVirtuesModeration2017}.

Alternatively, one can think of the costs more broadly as a form of \emph{friction} \cite{jiang2022trade}, perhaps by a specific task(s) that requires time and effort \cite{grimmelmannVirtuesModeration2017}. For example, in an attempt to reduce email spam, researchers recommended requiring email senders to compute a moderately expensive function before their message would be routed to a recipient \cite{dwork1992pricing}. An example more relevant to social media is Reddit’s decision to quarantine  problematic subreddits (e.g., r/the\_donald). Quarantined subreddits still function, but they are no longer publicly accessible and users must explicitly opt in to access the community \cite{jiang2020toward,chandrasekharan2022quarantined}. Platforms also employ a mix of both monetary cost and time or effort-based friction. An online community, MetaFilter, charged a \$5 entry fee and imposed a waiting period before new members could make new posts \cite[p. 200]{krautBuildingSuccessfulOnline2012}. In addition to raising a modest amount of revenue, the monetary cost and waiting period were acceptable to members who were a good match for the community, but not for those who might have disrupted it.

However, inducing friction may unintentionally discourage all user behavior, including legitimate user appeals or reports.
As a major example, research has shown that users found filing reports of online harassment on platforms to be burdensome and ineffective \cite{marwick2014online,mahar2018squadbox,blackwell2017classification}. 
% When cost structures induce only undesirable forms of participation occur, economists refer to it as adverse selection, as in a ``market for lemons'' where the only products that are sold are those that are of lowest quality \cite{akerlof1978market}. 
Thus, the design challenge here is to induce friction that can lead to a selection effect among users and deter only undesirable participation. In this work, we present a system that reasonably increases the effort cost of submitting a ban appeal. Appealing users must provide additional information for their appeals to be reviewed by moderators. However, for those with legitimate appeals the effort is not wasted; the process solicits information that moderators normally use to evaluate appeals, and successful users would have provided it anyway. Users with insincere appeals or those who want to send toxic messages to moderators are more likely to be discouraged by the effort and abandon their appeal. 

\begin{figure}
    \centering
    \includegraphics[width=0.6\textwidth]{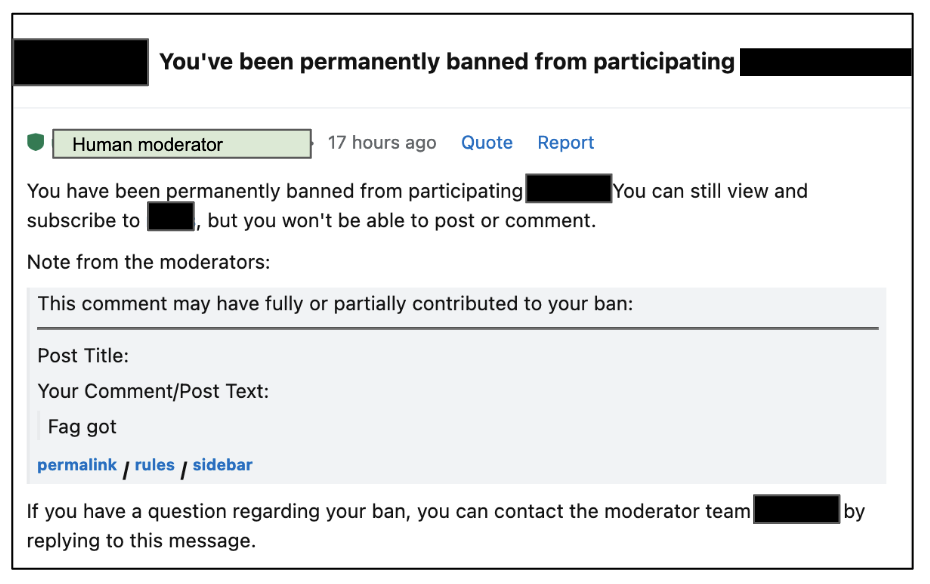} % first figure itself
    \caption{An example ban message shared with the banned user.}
    \label{fig:ban-message}
\end{figure}

 \begin{figure}
    \centering
    \includegraphics[width=0.6\textwidth]{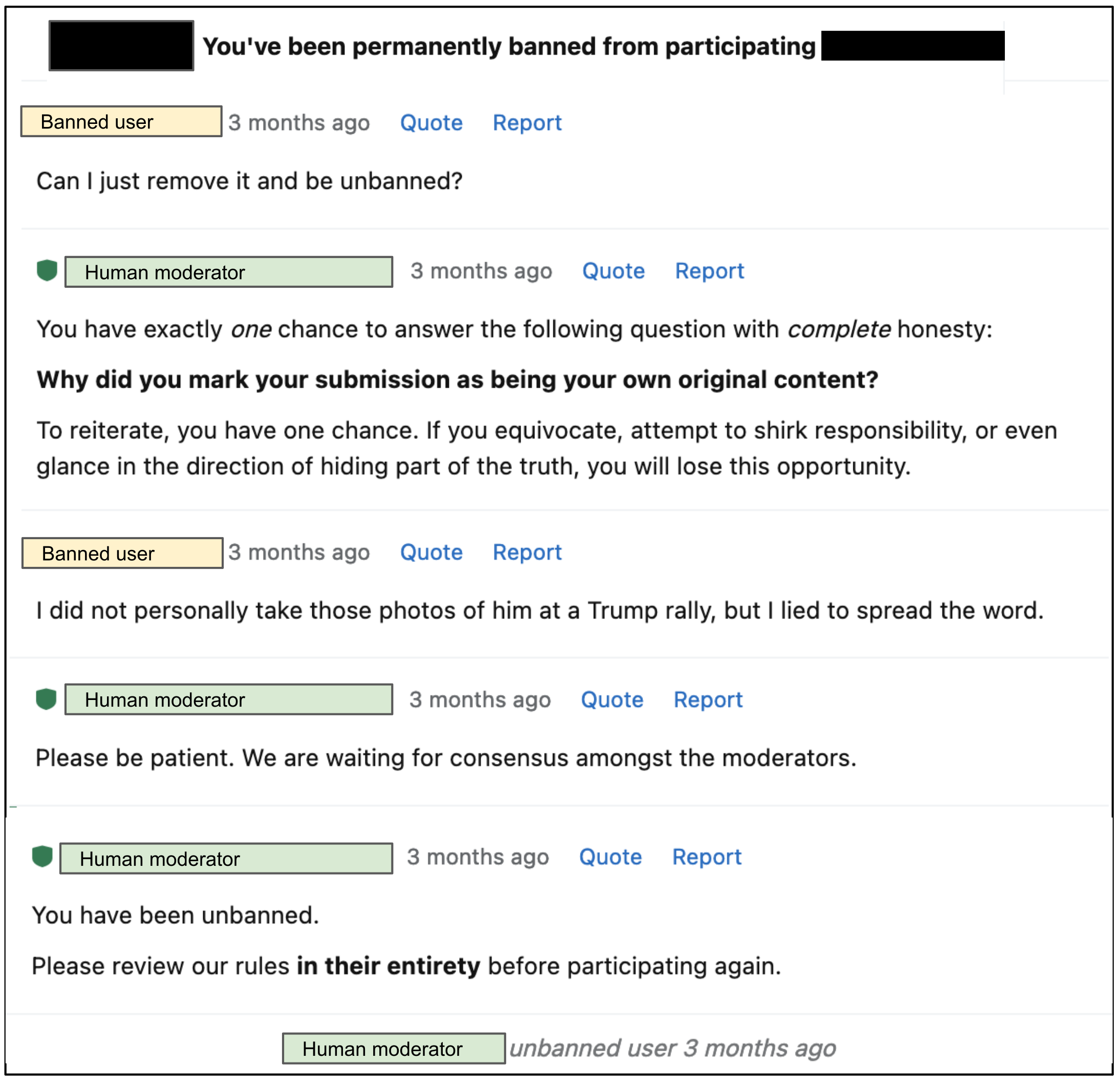} % second figure itself
    \caption{An example interaction between an appealing user and human moderators following the user's ban.}
    \label{fig:control-convo}
\end{figure}

\section{Study Context: Ban Appeals on Reddit}

In this section, we provide an overview of Reddit's current ban appeals process. On Reddit, moderators of individual communities are empowered to manage their own activities. Each individual community is called a subreddit. In terms of membership, moderators cannot control who joins their subreddit but they can ban users from participating in it. Moderators report banning users is an effective strategy for reducing spam and trolling behavior in their community \cite{seeringModeratorEngagementCommunity2019b}. According to Reddit's transparency report, over 4.9 million users were banned in 2021 across all subreddits \cite{redditincTransparencyReport}. 

When a user is banned from a subreddit, they receive a message from the moderators of that subreddit, informing them of their ban (and the ban reason) via modmail. Modmail is a shared messaging system on Reddit that moderators can use to communicate with individual members of their subreddit and vice versa. Figure \ref{fig:ban-message} shows an example ban message that was sent to one of the users from the study subreddit. The user was banned for intentionally misspelling the word fa***t to try and get around the subreddit's moderation bot which automatically deletes any comments containing that word. 

Reddit allows users to appeal against their ban by responding to their ban message. The first message sent by a user in response to their ban is automatically classified as their ban appeal. Ban appeals often lead to follow-up conversations between moderators and the banned user. Figure \ref{fig:control-convo} shows an example conversation between a banned user and moderators of our study subreddit. First, the banned user responds to their ban message to initiate their ban appeal process. The subreddit has a strict policy on when users can claim a picture to be their ``original content'', and the user was banned for breaching that policy. The moderator responds to the user's appeal by asking them to further explain their behavior. After some back and forth with the banned user, the moderators decide to grant their appeal.

Generally, moderators may adopt any of the following course of action in response to a user appeal -- 1) immediately deny or grant the appeal, 2) engage with the user in a follow-up conversation (and then deny or grant the appeal), 3) ignore and hide their appeal, or 4) mute the user and prevent them from sending any more messages for up to 28 days. 
Users may be banned temporarily (for a fixed period of time, ranging from weeks to a month) or permanently. For our study, we only consider permanent bans, as users are much more likely to appeal permanent bans due to their long-term consequences. While users can create multiple accounts on Reddit, using an alternate account to circumvent their ban is prohibited, and usually results in the user account being suspended from Reddit as a whole\footnote{\url{https://www.reddit.com/r/modnews/comments/wrnnvb/piloting_a_new_ban_evasion_tool/}}.  

We iteratively designed and built a system called AppealMod to support moderators' work on reviewing such ban appeals. We evaluated the system by deploying it in the subreddit called r/pics.  \add{r/pics is one of the largest communities on Reddit with over 29 million total members\footnote{The subreddit had roughly 29.8 million users as of January 2022}. It is the most popular place for sharing photos on Reddit and frequently features on Reddit's front page, which is visible to all Reddit users. The r/pics community is an attractive avenue for users who want to share their visual art with a broader audience; at the same time, its large membership and popularity also invites spam and other problematic behaviors. Overall, this results in a significant workload for the volunteers who are responsible for moderating the community. The moderators have drafted strict policies to govern the kind of content that is allowed in the community and often use automated tools to enforce some of these policies. The large number of moderation decisions that are made everyday (many of which are automated) invite further complaints and appeals from the users. Moderators' preference to manually review these complaints further adds to their workload. Through AppealMod, we aim to reduce the volunteer moderators' workload of reviewing user appeals.} All screenshots provided in the paper are actual conversations between users and moderators of r/pics. \add{Moderators provided permission to use the conversations in our paper.} The screenshots have been anonymized and edited for clarity.

%% file: content/design.tex
\section{Design Process and Moderator Needs}
% \jane{[Maybe ``Formative Study and Design Process'' is a more accurate section title. I think you mentioned being concerned about N=9, but I think it's okay as long as you emphasize the iterative design process and collaobration with the subreddit. The fact that you were able to continuously get mods' feedback is one of the  strengths of this work!]}

We used a collaborative and iterative design process to understand Reddit moderators' current practices of handling user ban appeals and design a system that addresses their needs. We began forging connections with Reddit moderators by interviewing nine moderators from different communities and inviting them for a future research collaboration. Then, we conducted three interviews to understand moderators' current practices and needs for handling user ban appeals. Finally, we focused on one community as a use-case and conducted design sessions with the moderation team to iterate over the design and evaluate it in a real world setting. The moderators generously shared their time, expertise, and ideas with us, and this deep, continuous engagement is a hallmark of our project. In the next two subsections, we describe our design process, provide an overview of the moderators' current practice of handling user ban appeals, and summarize their key needs that informed the design of our system.  

\subsection{Method: Collaborative and Iterative Design Process with Moderators}

\input{tables/participants}

\subsubsection{Building connections with moderators}
Fostering connections with community moderators is integral for developing a system that will be used by and deployed within the community. We started by interviewing nine moderators who collectively managed over 100 different subreddits. \add{We selected these moderators via purposive and snowball sampling. Given our emphasis on building partnerships with moderators, we customized recruitment messages and focused on communities where our first author could demonstrate commitment and interest. This allowed the first author to express their shared interests in the growth and welfare of the community while inviting the moderators to participate in the study. An example recruitment message is provided in the Appendix \ref{app:recruitment}}  

Details about our participants are provided in Table \ref{tab:participant-demographics}. We were broadly interested in understanding the major challenges moderators faced, and the interview scope was not restricted to their handling of user ban appeals. The first author, who is an active participant in several Reddit communities, conducted all interviews between April and July 2021. The interviews were conducted in English over Zoom and lasted for approximately 1 hour. During the interviews, moderators elaborated on the major issues they face, their coping strategies, and the scope for technology to support their work. At the end of the interviews, we asked moderators if they would be interested in a future research collaboration. Among the nine participants, three moderators expressed interest in a future collaboration and signed up for another round of interviews. These included moderators of r/pics, r/soccer, and r/politicalhumor. 

\subsubsection{Understanding moderators' handling of users ban appeals}

In the second round of interviews, we focused on the particular task of handling ban appeals on Reddit. Moderators during the first round repeatedly brought up their workload of addressing ban appeals from users. Furthermore, as we reviewed prior studies that focused on volunteer moderators, we found little evidence on understanding or supporting the moderators' current practices for addressing user ban appeals (see Section \ref{sec:background_tools} for a detailed review of prior work). All three moderators who signed up for a potential collaboration further expressed their interest in supporting the design of a moderation tool for addressing user ban appeals.

% Reddit moderators cannot control who joins their community, but they have the power to ban members. As a result, moderators actively used banning as a strategy to keep problematic users away from their community. Given the long-term consequences of a permanent ban, Reddit requires that moderators must allow users to appeal bans.

As we conducted a second round of three interviews with these three moderators, we centered the narrative around previous instances of ban appeals. For instance, when moderators described their potential approaches to address a ban appeal, we asked them to find examples of previous appeals that demonstrate this approach. We also asked follow-up questions about their use of technological tools and other resources during this process. Lastly, we asked them for any potential design ideas to reduce their workload. The interviews lasted between roughly an hour to an hour-and-a-half. All interviews were conducted in English over Zoom and transcribed by the first author.

We analyzed the interview data using \add{interpretive qualitative analysis \cite{merriam2002introduction}}\remove{qualitative} approach as outlined in \cite{jhaverDesigningWordFilter2022}. In particular, we began with open coding \cite{holton2007coding} to categorize our data into relevant patterns and then group them into appropriate themes. Next, we engaged
in multiple subsequent rounds of coding and memo-writing, conducting a continual comparison
of codes and associated data. The codes and emerging themes were discussed in weekly project meetings among coauthors. The first-level codes were specific, such as moderators looking for an apology, low effort appeals from users, etc. After several rounds of iteration, the codes were condensed into high-level themes, such as getting more information from users and effort asymmetry between moderators and users. Once authors agreed upon the themes, we used them to generate the design ideas that were used in the iterative design process. The themes are described in Section \ref{mod-needs-practices}.

\subsubsection{Iterative design process} \label{iterative-design-process}

During interviews, all moderators emphasized the need to solicit additional feedback from their community's entire moderation team before finalizing the design and evaluating it in a real setting. Therefore, for our iterative design process, we decided to focus on designing for one of the three communities, r/pics. r/pics is one of the largest subreddits with over 29 million total members. It is the most popular place for sharing photos on Reddit and frequently features on Reddit's front page, which is visible to all Reddit users. 

In September 2021, we shared a detailed proposal with all moderators of r/pics via modmail, introducing and explaining the design of our new tool and the evaluation process. The moderation team gave largely positive feedback and expressed interest in supporting the design and evaluation of the tool. At the time, r/pics had 8 moderators, and we invited all of them to participate in the design process. Between October and November 2021, we held two synchronous design sessions on Zoom. 
\add{The protocol we followed for these design sessions is provided in the Appendix \ref{app:design-protocol}.}
Three of the moderators attended these sessions, and each lasted for approximately one hour. In between these sessions, we also held a series of asynchronous discussions (over modmail) with the r/pics moderators. Five out of the eight moderators provided feedback during these discussions. For our design process, we used \add{multiple} low and mid-fidelity prototypes to present the design ideas generated from our earlier qualitative analysis. For instance, we \add{first used a flowchart to provide moderators with an overview of the process. The flowchart highlights that appealing users will receive automated messages in response to their requests and that some of the users' requests may remain hidden from moderators. A snapshot of the flowchart is provided in the Appendix (Figure \ref{fig:flowchart}). In addition to the flowchart, we also shared a potential typology for questions to be asked of the users during the process (see Figure \ref{fig:questions-type} in Appendix). Finally, to solicit additional feedback on our proposed automated bot, we provided moderators with a PowerPoint mockup conversation between appealing users and the bot (see Figure \ref{fig:prototype} in Appendix). These materials were sent to the moderators in advance and used during the design sessions. To keep this section concise, we included our design materials in the Appendix.} \remove{solicited moderator feedback on our proposed automated bot by presenting a PowerPoint mockup conversation between appealing users and the bot (see Figure \ref{fig:prototype} in Appendix).} 

All moderators were offered a compensation of USD \$20 (or an equivalent amount in their local currency) for each session or interview they participated in. One moderator respectfully declined the compensation. The entire design process was reviewed and determined to be exempt from oversight by our Institutional Review Board (IRB). 
% for camera-ready version: HUM00206815

\subsubsection{Pilot deployment}
\label{sec:pilot_deployment}
In January 202\remove{1}\add{2}, we carried out a pilot deployment of the tool for 3 weeks in r/pics under the complete supervision of its moderators. During this time, 88 banned users were subject to the AppealMod process. 34 out of 88 users completed the process. The research team analyzed the log data from the pilot deployment to resolve any bugs in the system, and we asked moderators for any feedback. The log data that we analyzed included 1) messages sent by the users and our bots' response to these messages, and 2) any messages or actions taken by the moderators, including their final decision. Moderators provided their feedback via modmail. 

% In the next subsection, we describe the findings on understanding the moderators current practices of handling ban appeals and the..TK TK

\subsection{Findings: Understanding Moderators' Practices and Needs}
\label{mod-needs-practices}

In this subsection, we report the findings from our design process. To provide our readers with a comprehensive understanding of moderators' current practices and their needs, we report the combined findings from the interviews \add{with mods of r/pics, r/soccer, and r/politicalhumor} and the design sessions \add{with r/pics mods} under relevant themes below. To help our readers contextualize the changes we made in our design after the pilot deployment, we summarize these changes in Section \ref{sec:appealmod_pilot} after describing the design features of our system.

\textbf{Reviewing users' past behavior:} When reviewing a user's appeal, moderators first looked for some evidence of positive behavior from the user. For example, they would try to find out if the user is active in other communities on Reddit and how they are behaving there. Regularly engaging in other communities was viewed favorably, whereas further instances of problematic behavior were detrimental to the user's appeal. As PIC\_M3 put it, \emph{``Are they a 4 or 5 year old account that is heavily active all over Reddit? Or is all of their history talking about Ivermectin and whether Covid vaccines are fake?''} \add{Reviewing a user's past behavior also allowed moderators to gauge the intent behind their behavior. According to POL\_M1: 
\begin{quote}
    ``I would say most of the time, especially with bans, I like to know what's going on with that account a little bit. One of the reasons is sarcasm as sometimes it is very hard to tell people being sarcastic, so [I] look at their account and see if they're a normal person, or if they're just a horrible person.''
\end{quote} 
In some cases, moderators also considered the norms of other communities a user is participating in as a useful signal in evaluating their appeals. As PIC\_M2 pointed out: 
\begin{quote}
    ``You can do the same thing if a racist comment is what they were banned for. There are certain subreddits that are known to be more racist, and if they have a lot of comments in those subreddits, then yeah, it's one indicator at least.''
\end{quote}
}

To support the moderators' review, Reddit allows them to directly access the information on a user's past behavior from their modmail interface. However, many banned users do not have such old accounts, or they may not be as active in other communities. Furthermore, Reddit allows users to delete their contribution history, making it impossible for the moderators to access that information.

\textbf{Getting more information from users:}  In case information on a user's past behavior was absent or insufficient, moderators reported directly engaging with the user to try and find out more about them. For example, when investigating the ban appeal in Figure \ref{fig:control-convo}, moderators were interested in finding out if the banned user was willing to accept responsibility for their actions. More generally, moderators reported asking users to explain why they think they were banned and then observed the user's reaction to their question.  

Moderators explained they were more likely to grant appeals from users who were apologetic or who put effort into explaining the circumstances that led to their behavior. Users who did not accept their mistake, blamed the moderators, or sent toxic messages to the moderators were less likely to have their appeals granted. \add{As PIC\_M2 pointed out about their interactions with banned users:
\begin{quote}
    ``They might be more polite about it, which could be built into a questionnaire workflow, or they might double down and be like, go f**k yourself, you f*****g snowflake, you know? And then that's just where the conversation ends.''
\end{quote}}

Moderators also gauged the user's awareness of their community norms, usually by asking them to read the rules and then explain whether or how they broke the rule(s). Moderators felt that users who demonstrated an awareness of the rules would be less likely to violate them again in future. 
\add{As noted by SOC\_M1:
\begin{quote}
    ``We usually ask direct and open ended questions like, why were you banned? Or why do you think you were banned? So you know, as I told you before, we want users to have self awareness but also have awareness about our rules.''
\end{quote}
} 
The overall approach was largely similar across all moderators and communities. In a few cases however, moderators took an approach that was more specific to a user's past behavior. For instance, if the user was banned for spreading misinformation, a moderator may ask the user if they were willing to remove their contributions and stop participating in communities that are known for discussing conspiracy theories.  

\textbf{Effort asymmetry and increased workload:} When reviewing ban appeals, moderators considered it crucial to carry out this investigative work given the long-term consequences of a permanent ban. However, the investigation process significantly increased their workload. In particular, it resulted in an \emph{effort asymmetry} between moderators and users -- users could submit their appeal with minimal effort while moderators bore the burden of this investigative work. For instance, in Figure \ref{fig:control-convo}, the user wrote a single sentence in their appeal; in response, moderators had to review their past behavior, ask further questions, and then make a final decision based on what they found. In a few cases, this led the moderators to explicitly ask for more effort from the user. As described by PIC\_M2, \emph{``I wanted something that shows they put effort into it. So I asked them to do something like write at least 200 words in their appeal. That would may be show that they are acting in good faith''}.  

% So there's only a limited time which we can invest into looking at modmail answering and unbanning. So if it isn't like a clear cut case, where we see immediately, either we did a mistake, or they did a mistake, and they're completely owning up to it, then it's just it like you would have to invest a lot of time to look at their profiles researched him to see okay, is it worth it or not? And it's usually easier okay. We have 27 million people on this subreddit, then this person won't be a part of it.

\textbf{Exposure to toxic messages:} Moderators also reported that doing their investigation increased the risk of being exposed to toxic messages from users. \add{As POL\_M1 described:
\begin{quote}
``They [banned users] are kind of like pseudo scientific and smart, and also superior sounding to everybody else. They convince a lot of people in their initial comments but then if you talk to them, at any level, they get super vitriolic and super angry and just kind of offensive to everybody, no matter what.''
\end{quote}
} 
Furthermore, users who were angry at moderators' decisions or those who were banned for prior toxic behavior tended to send more toxic messages in their appeal. While moderators temporarily muted such users to prevent them from sending new messages, the action was taken after that fact, i.e., once they had already seen the toxic message. \add{As PIC\_M2 described, \emph{``when you check what they were banned for, and it's just them dropping the N word everywhere, them saying terrible things and being a clear troll, it's like they're not worth my time...[They were] muted for a month.''}}

\textbf{Interacting with banned users:} Despite the added risk of being exposed to toxic messages, moderators believed it was important to directly communicate with a banned user. Sometimes, interacting with users revealed a mistake in the moderators' earlier decision. For example, moderators could have mistakenly banned a user if they were not aware of the full context behind their behavior or, as PIC\_M3 described, \emph{``simply did not get the joke''}. As POL\_M1 put it, moderators believed it was crucial to have \emph{``that human touch...it is just easier for people to correct the mistakes of people''.}

\textbf{Agency over final decision:} Moderators also wanted to retain their control over each and every decision. Moderators were concerned that automated decisions will result in many false positives. For instance, an automated system may incorrectly classify a genuine appeal as toxic and reject it. As PIC\_M2 pointed out: 
\begin{quote}
    ``You will get into false positives where they [user] might go, what's wrong with saying fa***t. We don't want to necessarily nuke that opportunity with that person because they're asking a question about that word. And we don't want the bot to ignore such appeals.''
\end{quote} 
Moderators were also concerned that automated decisions will encourage adversarial behavior from the users. For example, POL\_M1 commented: 
\begin{quote}``The weakness of the [automated] system I would say is once you do it enough, other people kind of figure out what's going on. And then you get that kind of planned response, once there's a way to know the answer, then the value of the investigation is diminished''.
\end{quote}

\textbf{Adding notes and archiving appeals:} To manage their workload and coordinate more effectively, moderators regularly added private notes to a user's appeal as a way to hold internal discussions or request another moderator for a review. These notes are not visible to the user and allow moderators to keep all information relevant to a user's appeal in one place.  \add{As POL\_M1 noted, 
\begin{quote}
    {``A majority of the bans I give out are three strikes and you're out kind of bans. We use that toolbox program to make notes on people's accounts. Either my co-mods or I would put a note on an account, saying, `Oh, this is a repeat offender of this kind of behavior' [...] so if it's an old account, and they have no notes in their account, that's a good sign, that means they've been a consistently good user.''}
\end{quote}} 
Upon completing their review of an appeal, moderators archive the appeal to declutter their modmail inbox and focus on active conversations with other users. 

\textbf{Summarizing Needs:} We identified several moderator needs to support their handling of user appeals. First and foremost, moderators wanted to reduce their own workload of reviewing ban appeals by addressing the effort asymmetry between them and the banned users (N1). To support their review of a user's appeal, moderators wanted to observe the user's reflection on their behavior that led to their ban (N2) and find out about the users' awareness of their community norms (N3). User's responses to these questions directly influenced their final decision. They also wanted more protection from direct toxic messages sent by angry users (N4). Finally, they wanted an opportunity to directly engage with the appealing user before making their final decision (N5). 

% \begin{itemize}
    
%     \item[N1:] Moderators want to gauge users' awareness of their community norms.
%     \item[N2:] Moderators want to elicit and observe users' reaction to their behavior that led to the ban.
%     \item[N3:] Moderators want to reduce their workload and effort asymmetry between mods and banned users. 
%     \item[N4:] Moderators want protection from potentially toxic messages.
    
%     \item[N5:] Moderators want to directly interact with users.
%     \item[N6:] Moderators want to have control over the final moderation decisions.

\section{AppealMod}
\label{sec:appealmod}
% \jane{[Same comment as the one I left in the Introduction. I recommend briefly describing the high-level idea of friction-based moderation first.]} 

Based on what we learned about moderators' needs from the design process, we built AppealMod, a system that helps moderators process appeals from banned users.
AppealMod's main goal is to reduce moderators' workload while maintaining their control over final moderation decisions. The system focuses on one type of moderation decision -- whether to maintain user bans -- though the framework should extend to other moderation decisions as well. 

AppealMod enforces an information collection process that users must complete before their appeals are sent to human moderators. The process requires that users answer questions about their past conduct and their understanding of the community's rules. This process is automated via a bot to reduce human moderators' workload. Completing the process requires that users interact with the bot and put in more effort (toward answering the questions) than the existing processing of submitting their appeal. 
The process, and the extra effort it requires, is designed to discourage some users from continuing their appeals, especially when the appeals are insincere. Appeals from users who abandon the process are not forwarded to moderators, meaning moderators never see incomplete appeals from users assigned to AppealMod. In the remainder of this section, we outline AppealMod's features and explain how they connect with the moderators' needs described in Section \ref{mod-needs-practices}, provide implementation details, and summarize the changes we made after the pilot deployment. 

% Informed by the moderators' goals arising from our design process, we built AppealMod, a system that reduces human workload of contestability for content moderation decisions while still maintaining human control over the final decision. 

% We designed AppealMod for one particular type of decision -- banning or suspending user accounts -- although the principle and framework should be extensible to other types of decisions as well. 

% AppealMod employs automated bots to enforce a screening process that must be completed by users before their appeals are reviewed by human moderators. 

% The screening process, which calls for additional effort from the users, includes questions about their past conduct and their understanding of the rules. The effort remains meaningful as the users' responses will add context to their appeal and support moderators' decision making. 

% Yet, a mediation via automated bots and the screening questionnaire likely discourages some of the appealing users, especially those with a spurious or bogus appeal. Appeals that do not complete the screening process are hidden from human moderators, effectively reducing the moderators' workload of reviewing these appeals. 

% We now provide an overview of the different features of our system, which are designed in line with the goals identified above. 

\subsection{Design Features}

\subsubsection{Automated bot}

To address the effort asymmetry between moderators and users (N1), we deploy an automated bot that guides users to complete the AppealMod process.
\add{The bot is necessary because Reddit's current appeals process allows banned users to send messages directly to the moderators’ inbox. Moderators wanted a way to keep banned users' messages out of their queue until the users had completed the process, and only with automation could that be accomplished.} The bot responds to any and all messages from banned users until they complete the process. In line with how Reddit classifies ban appeals, the first message from a banned user is automatically considered their ban appeal. The bot ignores any messages from users who are not banned from the community.

\begin{figure}
    \centering
    \includegraphics[width=\textwidth]{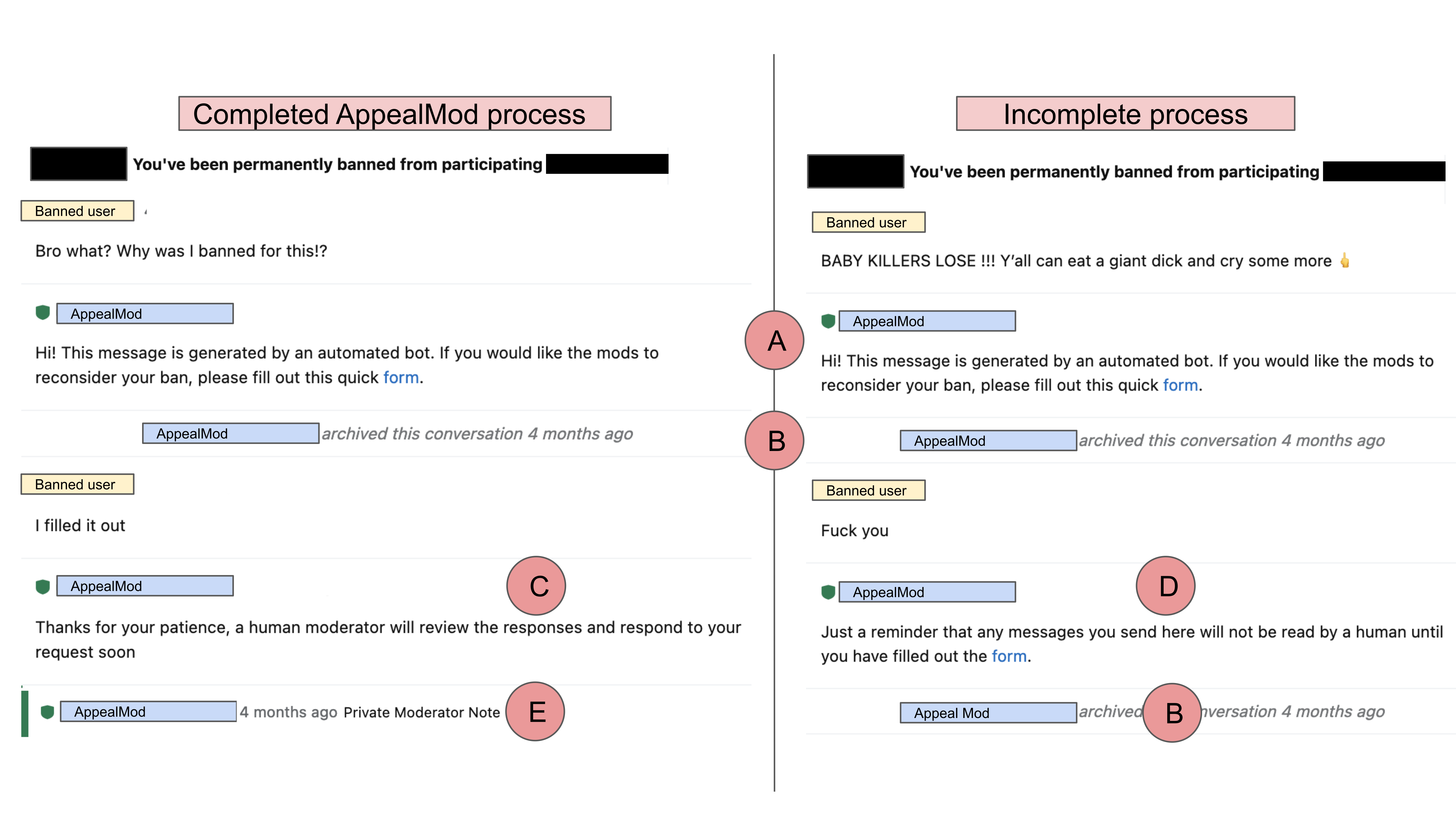}
    \caption{Two examples demonstrating a complete (left) and incomplete (right) AppealMod process. A) Our bot responds to the user's appeal and shares a link to the form containing our questions. B) Bot archives the conversation to hide it from human moderators. C) Once the user completes the AppealMod process, the bot hands over their appeal to human moderators. D) If the user sends any more messages before completing the process, they are reminded to complete the process. E) For users who complete the process, a private note summarizing their responses is also shared with human moderators.}
    \label{fig:treatment-convo}
\end{figure}

In response to a banned user's appeal, the bot asks them to complete the AppealMod process to have their appeals reviewed by human moderators (see Figure~\ref{fig:treatment-convo}-A). As part of the AppealMod process, the user has to answer questions (see Section \ref{sec:behavior-reflection} and \ref{sec:community-norms}) hosted on a webform. The webform can be easily updated and customized by the moderators without requiring any technical reconfiguration of the bot.
% In the future, the webform will allow moderators to easily update and customize these questions, without any technical reconfiguration of the bot. 
If the user sends any subsequent messages before completing the process, the bot reminds them to complete the AppealMod process before they can engage with human moderators (Figure~\ref{fig:treatment-convo}-D). Once the user completes the AppealMod process, they are informed that the appeal is handed over to human moderators for their review (Figure~\ref{fig:treatment-convo}-C).

% We describe the bot's dialogue flow and screening questionnaire in the subsection below. 

% These bots are necessitated on platforms that afford a direct communication channel between users and moderators (e.g., Reddit, Twitch, YouTube, etc.)

% Overall, through automated bots, AppealMod ensures that all users get the opportunity to present and have their cases reviewed by human moderators, without requiring any initial response from the moderators.   

\subsubsection{Hidden appeals}
While a user completes the AppealMod process (or simply ignores it), their appeals and any subsequent messages are archived and remain hidden from human moderators (Figure~\ref{fig:treatment-convo}-B). Archiving appeals is crucial for reducing the effort asymmetry (N1) so moderators do not have to put effort into reviewing incomplete user appeals. It also declutters their inbox and allows them to focus on currently active and important conversations with other users and moderators. Moderators regularly archived conversations that were no longer active as Reddit does not allow them to delete any conversations.

The default setting of hiding all messages from banned users also offers protection from toxic messages sent by angry banned users (N4). We expected these messages to remain hidden from human moderators, as users who intended to harass moderators by sending toxic messages were likely not willing to complete the AppealMod process. 

% While moderators have an additional option to access archived conversations from their interface, we do not expect them to spend their effort on going through these conversations. 

\subsubsection{Behavior reflection}
\label{sec:behavior-reflection}
To allow moderators to gauge a banned user's reflection on their behavior (N2), AppealMod asks two open-ended questions. First, the system asks them to describe their behavior and the circumstances that made them act that way. Then, it asks users to reflect on whether or how they might behave differently in the future. We designed these questions to be open-ended, as writing open-ended responses requires more effort and is a costly signal to replicate, compared to providing close-ended responses. For instance, it would be more difficult to forge regret in an open-text field compared to simply selecting the option that they regretted their behavior. We provide the questions verbatim in Table \ref{tab:task-questions} in Appendix.

\subsubsection{Awareness of Community Norms}
\label{sec:community-norms}

To gauge a banned user's awareness of community norms (N3), AppealMod first asks them to describe, in their own words, the rule that led to their ban. \add{By asking users to describe the rule they broke, the task nudges users to read and understand the rule that led to their ban---so that they can draft an appropriate appeal.}

\add{We also added a second task to evaluate the user's general awareness of the community norms. As part of this}\remove{We also added a} comment labeling task, \remove{in which} the user receives a set of five comments and must select comment(s) that they think should be allowed in the community. \add{This second task evaluates the user's general awareness of community norms to gauge whether they are likely to break additional rules if their appeals are granted and they are given a second chance.} We designed this as a labeling task so users do not have to read or memorize all the written rules but rather demonstrate a practical understanding of the community norms. All the comments were vetted by the moderators who participated in design sessions and they picked two of them as permissible. The set of comments is provided in Table \ref{tab:comment-label} in Appendix. 

\subsubsection{Moderator Handover} 
\begin{figure}
    \centering
    \includegraphics[width=0.8\textwidth]{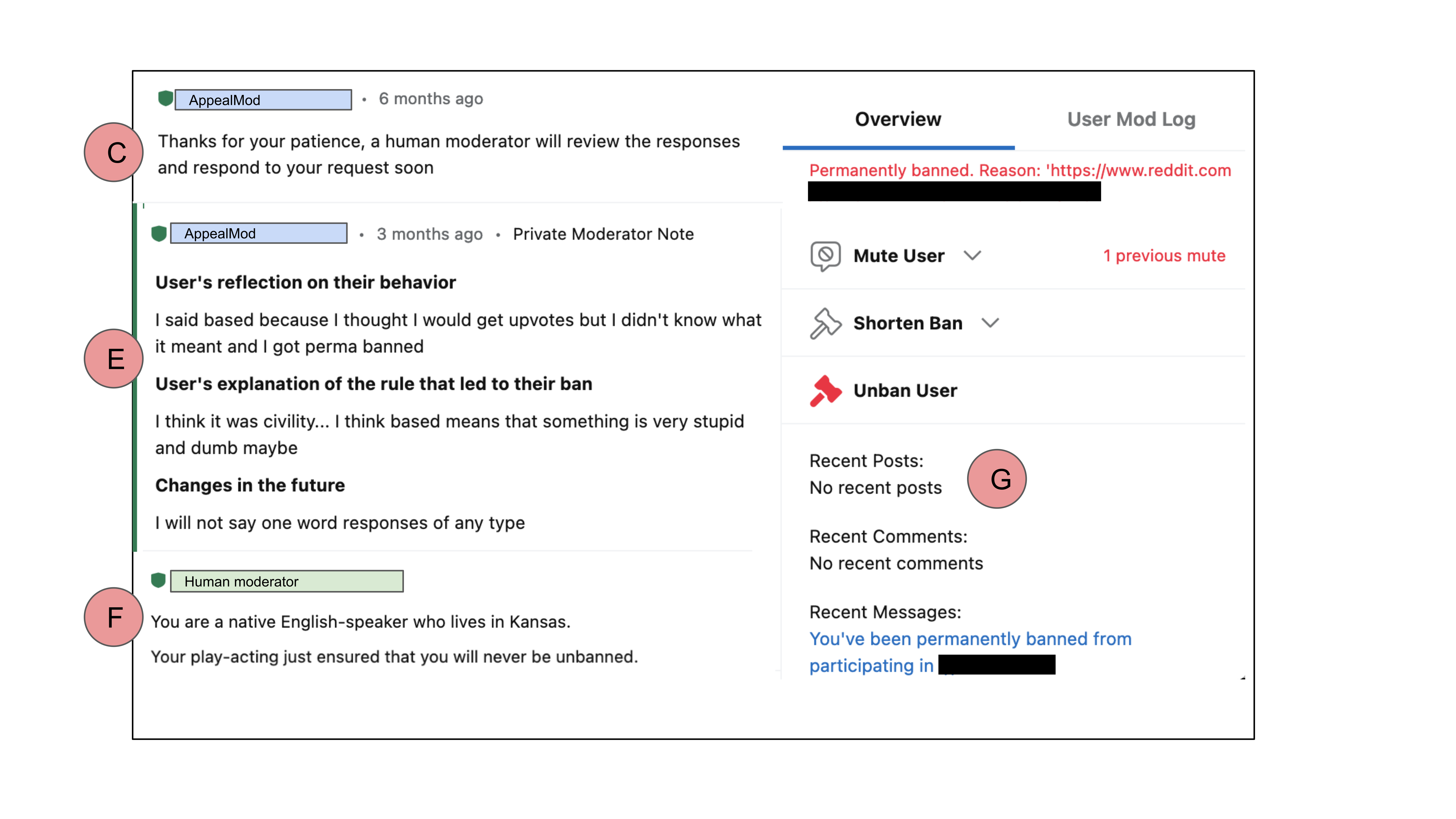}
    \caption{An example demonstrating how a user's responses are formatted and shared with human moderators (E). The information is readily available to moderators along with the user's past history already provided by Reddit (G). Moderators make the final decision by directly interacting with the user (F).}
    \label{fig:mod-note}
\end{figure}

Once a banned user completes the AppealMod process, their appeal is unarchived and appears at the top of moderators' inbox. This restores the direct communication channel between banned users and human moderators (N5). All the user messages are now visible to human moderators, and any subsequent messages in this conversation are ignored by our bot. The bot also adds a summary of the user's responses to this communication channel (see Figure~\ref{fig:treatment-convo}-E and Figure~\ref{fig:mod-note}-E). This additional information is visible only to moderators and is readily accessible when they review the user's appeal or engage with them further. The final decision on a user's appeal is always made by a human moderator (see Figure~\ref{fig:mod-note}-F).

\subsection{Implementation Details}
\begin{figure}
    \centering
    \includegraphics[width=\textwidth]{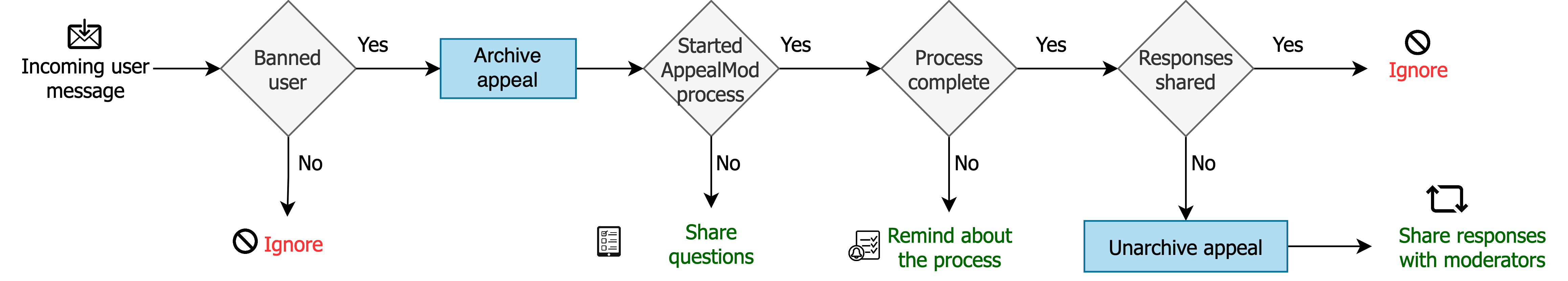}
    \caption{AppealMod Bot Dialogue Flow }
    \label{fig:dialogue-flow}
\end{figure}

The AppealMod bot is implemented in Python. The bot's responses are configured in a fixed dialogue flow, shown in Figure \ref{fig:dialogue-flow}. The dialogue flow is triggered in response to any new modmail message from a user. We used the Reddit API (with moderator-level privileges) so that the system continuously listens to any incoming messages from the users and performs other functions such as (un)archiving conversations, responding to users, and sharing private notes with moderators. During the experiment period, the bot ignored messages from banned users assigned to the control group. 

We stored information about a banned user's current state of appeal and their experimental group in a MongoDB database. The questionnaire was hosted via a Qualtrics webform. We used the Qualtrics API to dynamically control access to the webform, so that it is only accessible to banned users in the treatment group. A user could submit the form only once and could not change their responses afterward. We also used the Qualtrics API to periodically check whether a banned user had completed the AppealMod process, and if so, retrieved their responses. This way, we did not have to rely on any explicit confirmation from the user about completing their process.

\subsection{Changes After the Pilot Deployment}
\label{sec:appealmod_pilot}
The pilot deployment of the system in r/pics (described in Section \ref{sec:pilot_deployment}) also led to some crucial changes in the design of our system. For instance, we found in a small number of cases (4 out of 88) the ban reason was missing from the initial message sent to the user. Moderators attributed this to a bug in one of the moderation tools they use to ban users from their phone app instead of the website. Since the AppealMod process is designed to collect additional information concerning a user's ban, it would be meaningless to put users through the AppealMod process if they are unaware of their ban reason. Therefore, we updated our design to exclude those users from the AppealMod process so that their appeals are directly visible to human moderators who can notify them of the reason for being banned. 

Even when the ban reason was present in the initial message sent to the user, 11 out of 34 users who completed the AppealMod process responded to our questions saying they were unaware of their ban reason. In order to nudge users to read the ban message carefully, the first question in our webform asked them to copy and paste the ban reason provided in the initial ban message.   

We also made some changes based on moderators' feedback following the pilot deployment. Moderators mostly provided suggestions on improving the presentation of user responses collected during the AppealMod process. To incorporate their feedback, we used Markdown\footnote{\url{https://daringfireball.net/projects/markdown/}} style formatting to present the users' responses within the modmail interface (see Figure~\ref{fig:mod-note}-E). This improved the readability of the responses while making sure they were easily incorporated into the moderators' existing workflow.  

%% file: tables/participants.tex
\begin{table}[t!]

\caption{Details of moderators who participated in the design process}

\resizebox{1\textwidth}{!}{
\begin{tabular}{lcccccccccc}
 &  &  &  &  &  &  &   &    &   \multicolumn{1}{l}{\textbf{Participated in:}}    &                                                                      \\  \cmidrule{9-11}
\begin{tabular}[c]{@{}c@{}}\textbf{Participant}\\ \textbf{identifier}\end{tabular}     & \textbf{Subreddit}        & \begin{tabular}[c]{@{}c@{}}\textbf{Subreddit}\\ \textbf{topic}\end{tabular} & \begin{tabular}[c]{@{}c@{}}\textbf{Subreddit}\\ \textbf{size}\end{tabular} & \begin{tabular}[c]{@{}c@{}}\textbf{\# of subs}\\ \textbf{moderated}\end{tabular} & \begin{tabular}[c]{@{}r@{}}\textbf{Moderation} \\ \textbf{experience}\end{tabular} & \textbf{Gender} & \textbf{Country}    & \begin{tabular}[c]{@{}c@{}}\textbf{Building}\\ \textbf{connections}\end{tabular} & \begin{tabular}[c]{@{}c@{}}\textbf{Understanding}\\ \textbf{mods' needs}\end{tabular} & \begin{tabular}[c]{@{}c@{}}\textbf{Iterative} \\ \textbf{design sessions}\end{tabular} \\\midrule
P1 (SOC\_M1)  & r/soccer         & sports                                                    & 3.8 m                                                    & 1                                                                        & 6 yrs                                                         & M      & Argentina & x                                                              & x                                                                   &                                                                      \\
P2              & {[}redacted{]}   & hobbies                                                   & 19m                                                      & 12                                                                        & 3 yrs                                                         & F      & India     & x                                                              &                                                                     &                                                                      \\
P3 (POL\_M1)  & r/politicalhumor & satire                                                    & 1.5m                                                     & 56                                                                       & 10 yrs                                                        & M      & US        & x                                                              & x                                                                   &                                                                      \\
P4              & {[}redacted{]}   & lifestyle                                                 & 21m                                                      & 26                                                                       & 4 yrs                                                         & F      & US        & x                                                              &                                                                     &                                                                      \\
P5 (PIC\_M1)  & r/pics           & pictures                                                  & 29m                                                      & 26                                                                        & 11 yrs                                                        & M      & Scotland  & x                                                              & x                                                                   & x                                                                    \\
P6              & {[}redacted{]}   & history                                                   & 1.6m                                                     & 7                                                                       & 12 yrs                                                        & M      & Sweden    & x                                                              &                                                                     &                                                                      \\
P7              & {[}redacted{]}   & education                                                 & 40k                                                      & 1                                                                       & 9 yrs                                                         & M      & US        & x                                                              &                                                                     &                                                                      \\
P8              & {[}redacted{]}   & racial identity                                           & 5.6m                                                     & 24                                                                       & 8 yrs                                                         & M      & US        & x                                                              &                                                                     &                                                                      \\
P9              & {[}redacted{]}   & lifestyle                                                 & 8m                                                       & 21                                                                        & 3 yrs                                                         & F      & UK        & x                                                              &                                                                     &                                                                      \\
P10 (PIC\_M2) & r/pics           & pictures                                                  & 29m                                                      & 19                                                                        & 7 yrs                                                         & M      & Israel    &                                                                &                                                                     & x                                                                    \\
P11 (PIC\_M3) & r/pics           & pictures                                                  & 29m                                                      &  2                                                                      & 6 yrs                                                         & M      & UK        &                                                                &                                                                     & x                                                                    \\
P12 (PIC\_M4) & r/pics           & pictures                                                  & 29m                                                      & 34                                                                      & 8 yrs                                                         & M      & N/A         &                                                                &                                                                     & x                                                                    \\
P13 (PIC\_M5) & r/pics           & pictures                                                  & 29m                                                      & 42                                                                       & 7 yrs                                                         & M      & N/A         &                                                                &                                                                     & x                                                                   
\end{tabular}
}
\label{tab:participant-demographics}
\vspace{8px}
\end{table}

%% file: content/method.tex
\section{Evaluation} 
\label{evaluation}

We evaluated AppealMod by conducting a field experiment in r/pics. The experiment lasted 4 months, from April - August 2022. We and the moderators agreed it was necessary to deploy our system in a real setting to gauge the extent of effort users are willing to expend toward appealing against moderators' actions. 
% We wondered whether users were willing to put in the extra effort AppealMod required and if the system would adversely impact moderator workloads or decisions. 
The field experiment's main goals were closely tied to the moderators' needs described in Section~\ref{mod-needs-practices}. In line with the moderators' needs, we wanted to examine whether the system is effective at reducing moderator workload (N1) and protecting them from toxic appeals (N4). Additionally, we wanted to examine whether our system would adversely impact any appeal outcomes, such as the number of appeals granted by the moderators. From these goals, we identified six specific hypotheses tested in our experiment. We articulate those hypotheses below and then describe the experiment's design and outline our analysis plan.

% If we find that all users are willing to put that effort, the effectiveness of our system will be undermined as human moderators will roughly have the same amount of work to do. On the other hand, if very few users are willing to put that effort, it will reduce the number of appeals that are accepted, undermining the fundamentals of contesting moderation decisions. Consequently, we formulate our main hypotheses around the impact of our system on moderators' experiences as well as outcomes, such as appeal acceptance, that are relevant to contestability. 

% We now describe our experiment design in detail, followed by an articulation of our hypotheses, and the analysis we conducted to test these hypotheses. 

\subsection{Hypotheses}

\subsubsection{Impact on Moderator Workload}

We measure ``moderator workload'' as the number of appeals moderators receive and subsequently review. We expect that AppealMod will reduce the number of appeals that are visible to moderators because some users will be discouraged from putting additional effort and abandon their appeal rather than complete the process. 
% \footnote{From the point of view of moderators' workload, we are interested in the total number of appeals they respond to. The proportion of appeals that moderators respond to is analyzed separately as moderators' response rate under Hypothesis 3b stated below}. 
% Consequently, this would also reduce the total number of appeals that moderators have to respond to.
Therefore, our first hypothesis is:

\input{hypotheses/1a}

Since fewer appeals are visible to the moderators, the raw number of appeals they respond to will also reduce. Therefore, we predict hypotheses 1b:

\input{hypotheses/1b}

\subsubsection{Impact on Appeal Toxicity}

Users often criticize moderators for taking actions against them. As we found in our formative study, some users, instead of submitting sincere appeals, may send toxic messages in their appeals. We expect that users whose main intention is to attack or harass human moderators will be discouraged by the AppealMod process. As noted earlier, when users abandon the process, their messages remain hidden from human moderators. Therefore, our next hypothesis is:
%AppealMod effectively becomes a first-layer of defense for protecting human moderators from receiving toxic messages. 
 
\input{hypotheses/2a}

In contrast, users who complete the AppealMod process get the opportunity to directly engage with moderators. The additional effort AppealMod requires may frustrate these users and lead to more toxicity in their follow-up conversations with moderators. 
It is also possible that some users added toxic content to their form responses which are shared with the moderators.\footnote{Due to Perspective API's limitation, we did not automatically filter out user's form responses; see Section \ref{variables} for more details.} 
%Already having undergone the screening process, these users might find the overall process to be frustrating and cumbersome. Therefore, compared to the control condition where users directly interact with human moderators, AppealMod might increase the users' toxicity level in their follow up conversations with human moderators. 
Ideally, we want to ensure that AppealMod does not subject human moderators to additional toxicity from appealing users. To confirm, we propose the following hypothesis: 

\input{hypotheses/2b}

\subsubsection{Impact on Appeal Outcomes}

Our hypotheses so far have focused on evaluating the effectiveness of our system, AppealMod, at fulfilling moderators' needs, i.e., reducing their workload and protecting them from subsequent toxicity. It is also important to ensure that AppealMod doesn't adversely affect the desired outcomes on users' appeals. One measure to track is the number of appeals granted by moderators. Ideally, AppealMod will not impact the raw number of successful appeals; instead, it will reduce only the number of unsuccessful appeals that have to be reviewed by the moderators.
% If the underlying processes that determine whether a given user completes the screening process are completely random, then a decrease in the number of appeals forwarded to human moderators will proportionally reduce the number of appeals that are accepted by them. For instance, instead of AppealMod, if we implement a screening mechanism such that only 50\% of all appeals (randomly picked) are forwarded to human moderators, then we can expect the number of appeals accepted by moderators to also drop by roughly 50\%. 
%Alternatively, with AppealMod, we might expect some self-selection among users. 

Specifically, we expect that users who have a strong case for their appeal are more likely to self-select themselves into completing the process. In that case, a reduction in the number of appeals that are visible to human moderators would not lead to a decrease in the number of appeals that are granted. For our next hypothesis, we analyze the presence of self-selection among appealing users, and make the following claim:

\input{hypotheses/3a}

Another important outcome concerns with the human moderators' response rate on appeals and their follow up engagement with appealing users. Recent work on content moderation found that users of some social media platforms described the appeals process as ``speaking into a void'', due to its lack of human interaction \cite{myerswestCensoredSuspendedShadowbanned2018b}. Prior work also found that users considered communication, especially direct communication with human moderators, as one of the top three avenues for improving their abilities to contest content moderation decisions \cite{vaccaroContestabilityContentModeration2021a}. 

While AppealMod increases the cost of directly interacting with human moderators, it does not prohibit it. Users who complete the AppealMod process have a chance to interact with human moderators. Ideally, we want to ensure that AppealMod does not make it less likely that appealing users would receive a response from human moderators. In fact, we might even see an increase in moderators' responses given the extra effort initially put in by appealing users. %moderators would be prompted to respond given the effort already expended by users.  
We propose a conservative hypothesis to ensure appealing users still receive a human response when using AppealMod: 

\input{hypotheses/3b}

\subsection{Experiment Design}
We designed the experiment process in close collaboration with moderators of r/pics. Subreddit users were not made aware of the experiment. We took the decision as moderators believed that we will not be able to capture the users' true behavior if they were informed of the experiment. Furthermore, our system presented no more than minimal risk to the subjects. The experiment design, including a waiver of informed consent requirements, was approved by our Institutional Review Board.
\add{We describe our ethical considerations in more detail below.}
%University of Michigan IRB (HUM00206815).

During the experiment period, AppealMod managed appeals (and any subsequent messages) from banned users who were requesting reinstatement. AppealMod considered a banned user's first message to moderators via modmail as an appeal\footnote{Reddit uses the same criteria when displaying ban appeals as part of the moderators' modmail interface.} and randomly assigned the appealing user to either the \textit{control} or \textit{treatment} group. We maintained the user assignment throughout the experiment so that users who were banned more than once during the period would remain in the same group. However, none of the users whose appeals were granted were banned a second time.

Users assigned to the \textit{control} group experienced normal interactions as per Reddit's current ban appeals process. Their messages were immediately visible to moderators in modmail, and the AppealMod bot did not interact with these users. Figure \ref{fig:control-convo} shows an example interaction between an appealing user and a human moderator under control condition.

In the \textit{treatment} condition, users received automated messages from a bot that guided them through the AppealMod process. As we described in our system design (Section~\ref{sec:appealmod}), the AppealMod process required users to complete a webform. Until they completed the form, any messages they sent to the moderators were automatically archived and hidden from moderators. Once users completed the form, the bot added their responses to their modmail conversation, which was unarchived and made visible to human moderators. From this point onward, users experienced normal interactions with moderators analogous to the control group.

\paragraph{Ethical considerations}
\add{While we were not able to ask for the consent of users in r/pics due to the study design, we asked for and obtained moderators' consent as they represent the community and are knowledgeable about users' preferences. This is an approach used by prior work that involved deploying interventions in subreddits, such as \citet{zong2022bartleby}. The authors considered asking moderators' consent as asking for \textit{proxy consent} of the users in the subreddit, which is often used in social experiments when researchers cannot directly ask for participants' consent due to preserving the validity of the research \cite{humphreys2015reflections}. Instead, researchers disclose full information about the study to someone who can decide whether to give consent on the participant's behalf \cite{humphreys2015reflections}.} 

% \jane{[Mention debriefing with one mod (on behalf of the mod team)]}
%Control group represents the default behavior, where user appeals are directly forwarded to human moderators, who may choose to ignore the appeal or take subsequent actions (e.g., engage with the user or accept their appeal).  

%Under treatment, users are first subjected to the screening mechanism under AppealMod. As noted earlier, only appeals from users who complete the screening process are forwarded to the human moderators' inbox. All other appeals are archived and do not appear in the moderator inbox. 
%We set up the experiment so that a given user is always assigned to the same condition (i.e., either control or treatment) in case they are banned multiple times during the experiment period. However, for our experiment period, we found that none of the users whose appeals were accepted were banned again.

% As we have noted earlier, the system and the experiment were designed in close collaboration with moderators of our study subreddit. All users who appealed during the experiment period were eligible to be a part of the experiment. 
% Due to concerns of feasibility and validity, the appealing users were not informed of the experiment.  

% for some users (<10\%), the ban message did not contain the reason for their ban. These users were excluded from the experiment, as human intervention would be required to communicate the reason for their ban. 

% -- For all such cases, our bot left a note for the human moderators, nudging them to provide a reason for the user's ban. 

\subsection{Analysis}
\label{sec:analysis}

\input{tables/hypothesis}

Table~\ref{tab:hypothesis} summarizes the analyses we carried out for the hypotheses described earlier. Below we explain the additional variables computed and the statistical tests used to verify the hypotheses.

\subsubsection{Variables} \label{variables}
In addition to the outcome variables that can be directly captured from the experiment (e.g., number of messages sent, number of appeals visible, and whether an appeal was granted or not), we constructed two variables from the text of user appeal and messages: toxicity and predicted probability of success.

\textbf{Toxicity:} We measure human moderators' exposure to toxicity in the messages sent by users, and their responses as part of the AppealMod process. We use the Perspective API\footnote{\url{https://perspectiveapi.com/}} to assign a toxicity score to user appeals and any subsequent messages they sent. Perspective API has been used in prior work for measuring toxicity of texts \cite{mahar2018squadbox,im2020synthesized}. For any input text, Perspective API returns a score between 0 to 1, which denotes the likelihood of the message being toxic. \add{Among the scores provided by the API,\footnote{\url{https://developers.perspectiveapi.com/s/about-the-api-attributes-and-languages?language=en_US}} we used the TOXICITY score---which is described as measuring how likely the text is a ``rude, disrespectful, or unreasonable comment that is likely to make people leave a discussion.''}
We experimented with thresholds of 0.5, 0.7, and 0.9 to classify appeals and messages as toxic or non-toxic. For our main results, we set the threshold at 0.7. Results for other thresholds are reported in the Appendix.

We recognize that Perspective API is not robust to quoted sentences. For instance, consider the following text which received a toxicity score of 0.8 (i.e., 80\% likely to be toxic) -- \textit{``I said go fuck yourself because he was being disrespectful to women. Although, I accept that I should not have acted in anger.''} User appeals (and follow up conversations) often include similar quotations, and Perspective API classifies those messages as toxic. However, Perspective's scores in these cases should equally impact both control and treatment groups, and their comparison should be robust to this classification. 

Next, we also look for toxicity in the user responses provided during the AppealMod process. We manually analyzed these responses to only count instances of additional toxicity from users and excluded instances that had quotations to prior toxic behavior. We expected such responses from users since one of the questions asked them to reflect on their past behavior. For the manual analysis, one of the authors and one additional rater (a graduate student) independently rated a sample of 30 form responses. The rating guidelines defined toxicity as ``a rude, disrespectful, or unreasonable comment that is likely to make someone leave a discussion'' (as defined by the Perspective API). The raters were particularly instructed to look for instances of increased toxicity from users given their past behavior. The two raters met to discuss their disagreements and arrive at a consensus label. Each of the remaining form responses was then rated by one of the raters.

\textbf{Predicted Probability of Success:} 
Because AppealMod requires more effort, we expected some users to abandon their appeals. However, we didn't expect users to randomly abandon their appeals. Rather, we expected users whose appeals were likely to be granted to be more likely to complete the AppealMod process.
%However, we wanted to check for whether users who were likely to have a successful appeals will also abandon/complete 
%We expected that users who were more likely to be successful in their appeals were also more likely to do the extra work to complete the AppealMod process. 
To investigate this link, we wanted to predict an appeal's initial probability of success (PPS), which would be independent of whether the AppealMod process was completed or not. We modeled this likelihood of success using the initial appeal message, as the message was sent before the user's assignment to one of the experimental groups. To construct the PPS model, we collected (1) the initial message of all appeals made by users who were banned from r/pics between April 2021 and December 2021
(i.e., 6 months before the experiment began) and (2) the final decision on their appeal. The dataset consists of 6543 appeals, 846 (12.9\%) of which were granted. We use a simple Logistic Regression-based classifier. The classifier predicts the outcome of the appeal (granted or not) using unigram and bigram features from the initial appeal message. Using 5-fold cross validation, we found the model to be fairly accurate (F-score(macro)=0.83).

%To look for a potential self-selection effect among users who chose to complete (or skip) the screening process, we assign a prior probability of acceptance to their appeals. The prior acceptance probability is measured from the text of their appeal, which is drafted by the user prior to them being assigned to the treatment group (and subsequently being subject to the screening process). 

For all appeals under treatment during the experiment period, the output of this classifier was used as their PPS score. We then performed a logistic regression analysis to estimate if there was any relationship between PPS and whether users completed the AppealMod process. Based on our hypothesis, we expected users with higher PPS to be more likely to complete the AppealMod process. 

\subsubsection{Statistical analyses}
We used statistical tests to compare outcomes between control and treatment groups (see Table \ref{tab:hypothesis} for a summary). Specifically, we use a Chi-squared test to compare frequency or count variables (e.g., number of appeals granted) and \add{a Mann–Whitney–Wilcoxon test \cite{fay2010wilcoxon} to compare the}\remove{an independent samples t-test to compare the means of} distributions (e.g., number of messages exchanged between users and moderators per conversation). \add{Mann-Whitney-Wilcoxon test is a non-parametric test that does not require the samples to be drawn from a normal distribution \cite{fay2010wilcoxon}.} We also used regression analysis to characterize the relationship between the users' predicted probability of success (based on messages sent before they are assigned to an experiment group) and whether they completed the AppealMod process. %These test statistics are appended in the tables provided in the results section.

\subsection{Post-Experiment Session with Moderation Team}
Following the experiment, we had a debriefing session with one of the moderators of r/pics who participated on behalf of the entire team. In addition to quantitatively studying AppealMod's impact via the main study---a field experiment, we wanted to find out about the moderation team's perceptions of AppealMod. The first author conducted the session via Zoom which lasted for approximately 1 hour 30 minutes. We began by asking the moderator to describe their experience of addressing appeals via AppealMod. As the session progressed, we picked a random set of appeals from the treatment group and asked them to walk us through their process for addressing those appeals. \add{The complete protocol that guided the post-experiment session is provided in Appendix \ref{app:post-experiment-protocol}.} 
% mention the session protocol

%% file: hypotheses/1a.tex
\textbf{H1a: AppealMod will reduce the number of appeals that are visible to human moderators.}

%% file: hypotheses/1b.tex
\textbf{H1b: AppealMod will reduce the number of appeals that human moderators respond to.}

%% file: hypotheses/2a.tex
\textbf{H2a: AppealMod will reduce the proportion of appeals that are visible to human moderators that contain a toxic message.}

%% file: hypotheses/2b.tex
\textbf{H2b: AppealMod will not increase toxicity in users' subsequent engagement with human moderators.}

%% file: hypotheses/3a.tex
\textbf{H3a: AppealMod will not reduce the number of appeals granted by human moderators.}

%% file: hypotheses/3b.tex
\textbf{H3b: AppealMod will not reduce human moderators' engagement on appeals.}

%% file: tables/hypothesis.tex
\begin{table}[]
\resizebox{0.9\textwidth}{!}{
\begin{tabular}{l|l|l|l}

                                    & \textbf{Hypothesis}           & \textbf{Measure}                                                     & \textbf{Statistical test}           \\ \hline
\multirow{2}{*}{Moderator workload} & H1a                  & No. of appeals that are visible to moderators               & Chi-square test            \\ \cline{2-4} 
                                    & H1b                  & No. of appeals that moderators respond to                   & Chi-square test            \\ \hline
\multirow{2}{*}{Appeal Toxicity}    & H2a                  & Proportion of visible appeals that are toxic                & Chi-square test            \\ \cline{2-4} 
                                    & H2b                  & User toxicity in subsequent messages to the moderators      & Independent t-test         \\ \hline
\multirow{5}{*}{Appeal outcomes}    & \multirow{2}{*}{H3a} & No. of appeals granted                                   & Chi-square test            \\ \cline{3-4} 
                                    &                      & AppealMod completion $\sim$ predicted probability of success & Regression Analysis* \\ \cline{2-4} 
                                    & \multirow{3}{*}{H3b} & Ratio of appeals responded to and appeals visible           & Chi-square test            \\ \cline{3-4} 
                                    &                      & No. of messages exchanged                                   & Independent t-test         \\ \cline{3-4} 
                                    &                      & No. of characters exchanged                                 & Independent t-test         \\ 
\end{tabular}
}
\vspace{5px}
\caption{Overview of our hypotheses and their associated measures. *All statistical tests but the regression analysis compared between the control and treatment groups. Regression analysis was only applied to the treatment group.}
\label{tab:hypothesis}
\end{table}

%% file: content/results.tex
\section{Results}

We conducted the experiment for four months, from April to August 2022. During this period, 880 users appealed their bans. Users were roughly equally divided into control (438) and treatment groups (442). Out of the 442 users assigned to the treatment group, 131 completed the AppealMod process. The median time for completing the process (after clicking on the form link) was 4 minutes and 50 seconds. 90 out of the 880 appeals were granted by moderators. In the following subsections, we describe the results specific to our individual hypotheses. 

\subsection{Moderator Workload was Lower}

\input{hypotheses/1a}
\noindent\input{hypotheses/1b}

Table \ref{tab:workload} shows a comparison between number of appeals that were visible to human moderators and that they responded to under control and treatment. 
In the control condition, all appeals submitted (n = 438) were immediately visible to moderators. In the treatment condition, only 30\% of appeals (n = 131) were visible to moderators. Users abandoned the other 70\% of appeals (n = 311) in the treatment condition, and those incomplete appeals were not visible to moderators. Therefore, AppealMod reduced the human moderators' review queue by 70\%. Human moderators did not respond to all appeals in their queue under either condition. In the control condition, moderators responded to 263 appeals, and in the treatment condition, they responded to 105.

Chi-squared tests confirmed that fewer appeals were visible to the moderators in the treatment condition ($\chi^2(1, N=880) = 437.55, p=\num{5.39e-105}$) and that they responded to fewer appeals in the treatment condition ($\chi^2(1, N=880) = 117.59, p=\num{2.13e-27}$).

% Under control, all 438 appeals were directly forwarded to (and potentially read by) human moderators, whereas under treatment, only 131 (out of 442) appeals were forwarded to human moderators. That is, our system AppealMod reduced the moderators' workload of reviewing user appeals to almost 30\% (compared to the control group) as 70\% (311 out of 442) of the appealing users did not complete the screening process. Consequently, this also reduced the number of appeals that moderators responded to (263 in control; 105 in treatment). A Chi-squared test verified that both these differences were statistically significant (see Table \ref{tab:workload}). 

\input{tables/workload}

\subsection{Fewer Toxic Appeals were Visible to Moderators}

\begin{figure}
    \centering
    \includegraphics[width=0.4\textwidth]{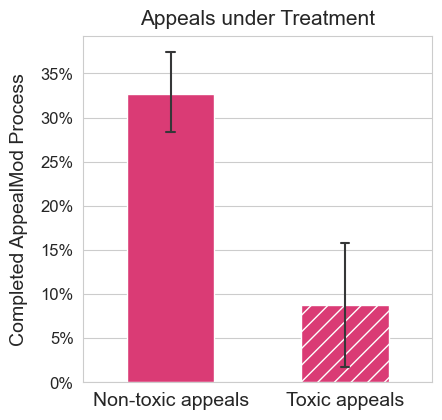}
    \caption{Comparing the proportion of toxic and non-toxic appeals with AppealMod process completed. \add{The 95\% confidence intervals for the two groups are non-overlapping, which indicates that these differences are large enough to be reliable.}}
    \label{fig:toxicity-self-selection}
\end{figure}

\input{hypotheses/2a}
Roughly 13\% (n = 115) of all initial user appeals were classified as toxic; with nearly equal numbers of toxic appeals in control (58) and treatment (57). Under control, all toxic appeals were directly visible to human moderators. However, under treatment, only 3.8\% of the appeals visible to human moderators were toxic (see Table \ref{tab:workload}); the other toxic appeals were hidden because users abandoned those appeals. A Chi-squared test verified that the difference in the proportion of appeals visible to human moderators that were toxic under control and treatment was statistically significant, $\chi^2(1, N=569) = 8.17, p=0.004$. We set a threshold of 0.7 on the Perspective API output to classify appeals as toxic or not. Varying the threshold to 0.5 or 0.9 did not qualitatively change the results (see Table~\ref{tab:appendix-toxicity-thresh} in Appendix). 

These results indicate that, under treatment, most users who authored toxic appeals did not complete the AppealMod process, and therefore, these toxic appeals were hidden from human moderators. Figure \ref{fig:toxicity-self-selection} shows that while 32.7\% of non-toxic appeals were completed, the completion rate dropped to 8.7\% for toxic appeals. Furthermore, the 95\% confidence intervals for the two groups are non-overlapping, indicating that these differences are large enough to be reliable.

\input{hypotheses/2b}
% \begin{figure}
%     \centering
%     \includegraphics[width=0.6\textwidth]{figures/follow-up-toxicity.png}
%     \caption{Comparing users' toxicity levels in their follow up conversations with human moderators under control and treatment.}
%     \label{fig:follow-up-toxicity}
% \end{figure}

For this hypothesis, we are interested in (1) comparing toxic messages sent by users under control and treatment when directly interacting with moderators, and (2) toxic content provided by treatment users in their form responses. For (1) we are only interested in appeals that had at least one back-and-forth interaction between the user and the moderators. In these cases, both the user and the moderator(s) sent at least one message each after the user's initial appeal. There were 162 conversations under control and 84 under treatment. Under control, users sent a total of 326 messages out of which 12 were classified as toxic. Under treatment, users sent a total of 207 messages out of which 10 were classified as toxic. A Chi-squared test verified that the difference in the proportion of messages sent by users that were toxic under control and treatment was not statistically significant, $\chi^2(1, N=533) = 1.57, p=0.209$. 

% Figure \ref{fig:follow-up-toxicity} shows the distribution of toxicity scores of follow-up messages sent by users to human moderators under both conditions. As mentioned in the analysis (Section \ref{sec:analysis}), toxicity score of a text lies between 0 and 1, where 1 indicates a high likelihood of the text being toxic. While the figure shows that treatment group had a higher proportion of messages with very low toxicity scores (< 0.1), the overall distribution of toxicity in messages sent under control and treatment is very similar. An independent samples t-test verified that the difference between the two distributions (i.e, control and treatment) is not statistically significant $ (0.87, p=0.38) $. In back-and-forth conversations, we did not observe differences between control and treatment in the distributions of users' messages' toxicity.  

Next, we look at the number of form responses from treatment users that were toxic. As we noted in Section \ref{variables}, we analyzed the form responses manually to count instances of additional toxicity from users and exclude instances that had quotations to prior toxic behavior. Out of 131 form responses, 3 were flagged as toxic. However, the rate of toxicity in form responses (2.3\%) is still lower than the base rate of toxicity among initial appeal messages (13\%). 

\subsection{Moderators Granted Similar Number of Appeals and had a Higher Response Rate}

\input{tables/outcomes}

\input{tables/regression}

\input{hypotheses/3a}
\begin{figure}
    \centering
    \includegraphics[width=0.4\textwidth]{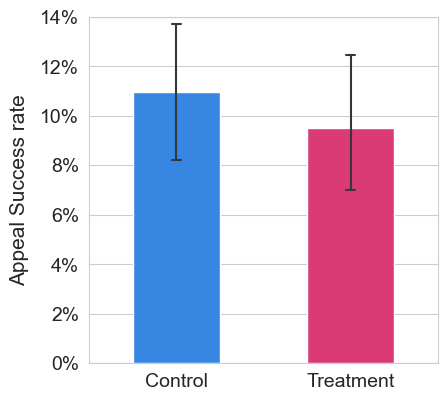}
    \caption{Comparing success rate of appeals under control and treatment. \add{The confidence intervals largely overlap, and a Chi-squared test further showed that the difference between the two are not statistically significant ($\chi^2(1, N=880) = 0.362, p=0.547$).}}
    \label{fig:acceptance}
\end{figure}

During our experiment, the overall success rate of appeals was roughly 10\% (90 out of 880). Figure \ref{fig:acceptance} shows the success rate of appeals and their 95\% confidence intervals under control and treatment. While 48 appeals were granted under control and 42 were granted under treatment, the large overlap between the confidence intervals suggests that success rates with and without the AppealMod process are similar. A Chi-squared test further verified that differences in these two results are not statistically significant ($\chi^2(1, N=880) = 0.362, p=0.547$).  

To further investigate the selection effect among users under treatment, we used a logistic regression model to analyze the relationship between users' predicted probability of success (PPS) and whether they completed the AppealMod process. As noted in Section \ref{sec:analysis}, a user's PPS can be reliably computed from the text of their initial appeal message (posted before they are asked to complete the AppealMod process). Table \ref{tab:reg-prior-prob} summarizes the regression model and shows a statistically significant $p<0.001$ relationship between a user's PPS and whether they completed the AppealMod process. Computing the odds ratio, we found that a 10\% increase in the users' PPS increased their odds of completing the AppealMod process by 40\%. 

These results indicate that the AppealMod process did not uniformly discourage users from continuing their appeals. Instead, users whom we predicted to have a high probability of success, based on their initial message, were more likely to complete the AppealMod process. Users with a low likelihood of success were less likely to complete the process. 

% This observation that users with higher prior acceptance probability are more likely to complete the screening process offers a justification for why the number of appeals accepted are largely similar under control and treatment. The requirement of a screening mechanism did not discourage all users uniformly; most users whose appeals were likely to be accepted still ended up completing the screening process. 

% \textbf{H3b: Users with higher prior probability of appeal acceptance are more likely to complete the screening process.}

\input{hypotheses/3b}
Table \ref{tab:outcomes} shows a comparison between the various outcomes on appeals that are visible to human moderators under control and treatment. First, we found that moderators' response rate (a desired outcome) on visible appeals under treatment (80.15\%) was higher than their response rate on visible appeals under control (60.04\%). A Chi-square test verified that the difference was statistically significant, $\chi^2(1, N=569) = 16.97, p=\num{3.78e-5}$.  

Next, we analyzed follow up conversations on appeals that moderators responded to. \add{Using a Mann–Whitney–Wilcoxon test,} we found no statistically significant difference between control and treatment groups in terms of the total number of messages exchanged on a given appeal, the number of messages moderators sent, or the length of moderators' messages. \remove{An independent samples t-test was used to verify whether these differences were statistically significant.} Under both control and treatment, a median of 5 messages were exchanged between the appealing user and moderators \add{(Mann–Whitney U = 6701.5, n=246, p=0.422, two-tailed)}\remove{($t(246) = 0.02, p=0.982$)}, and moderators sent a median of 2 messages per conversation \add{(Mann–Whitney U = 6108.5, n=246, p=0.085, two-tailed)}\remove{($t(246) = 1.51, p=0.132$)}. In terms of characters, moderators' contributed a median of 210 characters per conversation in control and 208 characters per conversation in treatment \add{(Mann–Whitney U = 6732.5, n=246, p=0.447, two-tailed)}\remove{($t(246) = -0.72, p=0.474$)}. 

\subsection{Notes from the Post-Experiment Session with Moderators}

We briefly note the findings from the post-experiment session with a moderator representing the team. Overall, the team was satisfied with AppealMod. The moderators confirmed that with the additional information provided during the AppealMod process, they found it easier to respond to users' appeals. \add{Specifically, moderators reported that users' responses gathered via the AppealMod process provided valuable information and context that is hard to get from the platform's existing signals. For example, the moderator said:
\begin{quote}``And it gives us the context because what we have to work with is very little. If they're talking a lot about their personal life in their post history, great, but the majority of Reddit accounts are not posting personal information.''
\end{quote}
In some cases, the users' responses also helped in gauging their awareness of the community norms. As the moderator put it: 
\begin{quote}
    ``So obviously, if they went back to doing exactly the same thing, they would get banned again. But the form tells me that they are aware of what our guidelines are, and they're stating that they're willing to abide by them. And that, to me, is enough for someone to be unbanned.''
\end{quote}
}

\add{According to the moderators, having an automated bot also streamlined the appeals process, making conversations with the user faster and more efficient. For instance, one moderator said: \begin{quote}
    ``When we didn't have the bot and in cases when it doesn't fire, we do all the investigation that we have to do. And when we have a conversation, we have to wait for the user to respond, and go back and forth several times. So that process often slows it down for the user -- if they miss a message and drop off, obviously we wouldn't press it. So having that [AppealMod] form both enables and speeds along the process.''
\end{quote}
}

\add{While the moderators found users' responses collected during the AppealMod process to be helpful, these responses alone did not determine the final outcome. Moderators followed a nuanced decision making process that, in addition to utilizing the AppealMod process, also involved additional factors, such as, the original reason behind a ban, who issued the ban, how old is the ban, and how the user behaving in other communities after their ban, to name a few. A moderator provided a relevant example: 
\begin{quote}
    ``Another contributing factor of why I banned them is when you look at their profile, they have negative comment karma, which is very difficult to do just because of the way that Reddit adds karma, which says to me that they are explicitly trolling and that's probably the goal of the account; they are not engaging in good faith.''
\end{quote}
}.

\remove{In most cases, it immediately prompted the moderators to ask more questions from the user. In a few cases, the responses directly encouraged moderators to give the user another chance. Generally, moderators were favorable towards users who wrote longer responses. The length and quality of a user's responses was, however, only one of the factors moderators considered in their decision making process. They still accounted for the user's behavior in other communities and their initial ban reason when making the decision.}

The moderation team also requested one change in the AppealMod design. Currently, the bot is configured to update moderators via a private note when the Qualtrics API is down and users' responses cannot be retrieved. This happened twice during the experiment period. The bot was able to retrieve responses once the API became functional after a few hours. Moderators gave feedback that users should also receive a message from the bot informing them of any technical issues and requesting them to be patient while the issue gets resolved.

The moderation team requested that we configure AppealMod for addressing all user ban appeals moving forward, removing the control condition. We have not yet done so, in part because we do not yet have a maintenance and monitoring plan that would allow us to provide the service reliably enough. With the current setup, where AppealMod is turned on for only some of the appeals, moderators are able to handle downtime of the system without disruption to their normal practices.

% The first author transcribed the session and coded 
% While the experiment results are the main contribution of this work, we briefly summarize how r/pics moderation team perceived AppealMod after having used it during the experiment, as well as their feedback on further improving it.
% \jane{[Brief summary; just one paragraph?]}

%% file: tables/workload.tex
\begin{table}[]
\resizebox{0.6\textwidth}{!}{
\begin{tabular}{p{0.25\textwidth}|c | c}

                                     & \textbf{Control}                 & \textbf{Treatment} \\ \hline

\add{Total appeals submitted} & \add{438} & \add{442}    \\ \hline

Appeals visible              & 438                     & 131                  \\ \hline
Appeals responded to                 & 263                     & 105                   \\ \hline
Proportion of visible appeals that are toxic & 13.24\% (58 out of 438) & 3.8\% (5 out of 131)           \\ 
\end{tabular}
}
\vspace{7px}
\caption{Comparison between appeals visible to human moderators under control and treatment}
\label{tab:workload}
\end{table}

% \begin{table}[]
% \resizebox{\textwidth}{!}{
% \begin{tabular}{|p{0.25\textwidth}|p{0.18\textwidth}|p{0.18\textwidth}|p{0.4\textwidth}|}
% \hline
%                                      & Control                 & Treatment            & Chi-squared test                               \\ \hline
% Total appeals forwarded                     & 438                     & 131                  & $\chi^2(1, N=880) = 437.55, p=\num{5.39e-105}$ \\ \hline
% Appeals responded to                 & 263                     & 105                  & $\chi^2(1, N=880) = 117.59, p=\num{2.13e-27}$  \\ \hline
% Proportion of forwarded appeals that are toxic & 13.24\% (58 out of 438) & 3.8\% (5 out of 131) & $\chi^2(1, N=569) = 8.17, p=0.004$             \\ \hline
% \end{tabular}
% }
% \caption{Comparison between appeals forwarded to human moderators under control and treatment}
% \label{tab:workload}
% \end{table}

%% file: tables/outcomes.tex
\begin{table}[]
\resizebox{0.75\textwidth}{!}{
\begin{tabular}{p{0.6\textwidth}|c|c}

                                     & \textbf{Control}                 & \textbf{Treatment}                 \\ \hline

Appeals visible & 438 & 131    \\ \hline

Moderator response rate on visible appeals  & 60.04\% & 80.15\%         \\ \hline
Median number of messages exchanged per appeal & 5 & 5        \\ \hline
Median number of messages sent by human moderators per appeal & 2 & 2          \\ \hline

Median number of characters contributed by human moderators per appeal  & 210 & 208       \\ 
\end{tabular}
}
\vspace{5px}
\caption{Human moderators' engagement on appeals that were visible under control and treatment}
\label{tab:outcomes}
\end{table}

% \begin{table}[]
% \resizebox{\textwidth}{!}{
% \begin{tabular}{|p{0.3\textwidth}|p{0.18\textwidth}|p{0.18\textwidth}|p{0.33\textwidth}|}
% \hline
%                                      & Control                 & Treatment            & Statistical test                               \\ \hline
% Appeals accepted & 48 & 42 & $\chi^2(1, N=880) = 0.362, p=0.547$             \\ \hline
% Moderator response rate & 60.04\% (263 out of 438) & 80.15\% (105 out of 131) & $\chi^2(1, N=569) = 16.97, p=\num{3.78e-5}$             \\ \hline
% Median number of messages exchanged per appeal & 5 & 5 & $t(246) = 0.02, p=0.982$             \\ \hline
% Median number of messages sent by human moderators per appeal & 2 & 2 & $t(246) = 1.51, p=0.132$             \\ \hline

% Median number of characters contributed per appeal  & 210 & 208 & $t(246) = -0.72, p=0.474$       \\ \hline
% \end{tabular}
% }
% \caption{Comparison between outcomes on appeals forwarded to human moderators under control and treatment}
% \label{tab:outcomes}
% \end{table}

%% file: tables/regression.tex
\begin{table}[!htbp]
\centering 
\resizebox{0.56\textwidth}{!}{
\begin{tabular}{@{\extracolsep{5pt}}lc} \\[-1.8ex]\hline \hline \\[-1.8ex] & \multicolumn{1}{c}{\textit{Dep Variable: AppealMod process complete }} \ \cr \cline{1-2} \hline \\[-1.8ex]  Intercept & -1.051$^{***}$ \\   & (0.117) \\  PPS & 3.408$^{***}$ \\   & (0.939) \\ \hline \\[-1.8ex]  AIC & 525.632 \\  Observations & 442 \\ Residual Std. Error & 1.000(df = 440)  \\  F Statistic & $^{}$ (df = 1.0; 440.0) \\ \hline \hline \\[-1.8ex] \textit{Note:} & \multicolumn{1}{r}{$^{*}$p$<$0.05; $^{**}$p$<$0.01; $^{***}$p$<$0.001} \\ 
\end{tabular} 
}
\vspace{10px}
\caption{Regression analysis on the relationship between predicted probability of success (PPS) and whether the AppealMod process was completed}
\label{tab:reg-prior-prob}
\end{table}

%% file: content/limitations.tex
\section{Limitations}

While the novel system we developed and experimented with shows promise for reducing moderator workload without negatively impacting the subreddit community, our work has several limitations worth noting. 

First, our system might lead to other unintended negative outcomes that we have not measured. We note that users who post toxic comments and users with insincere appeals were more likely to be discouraged by the AppealMod process. \add{However, some of the users discouraged from completing the AppealMod process did not display any toxic behavior. It is possible they indulged in other kinds of problematic behavior, for example, spamming bots will not be able to complete the AppealMod process, as it requires human intervention to manually fill out the form. However, since we did not collect any additional information from the users (such as their demographic information), we cannot ascertain whether any other factors played a role in users abandoning the process.}\remove{However, we cannot ascertain whether certain systemic factors, such as user demographics, also played a role.} For instance, we designed the AppealMod questionnaire in English as contributions to r/pics are limited to English language. It is possible that users whose first language is not English found it more challenging to respond to the questionnaire and were therefore more likely to abandon their appeal. More broadly, prior research has found that content moderation interventions can disproportionately harm marginalized users \cite{haimson2021disproportionate}. We encourage future work to explore whether our specific design of AppealMod and the more generic approach of inducing friction in moderation processes \add{has differential impact on different groups of users, and whether it} causes further harm to marginalized users.

Second, we found that AppealMod reduced the number of appeals reviewed by moderators by 70\%; however, we  evaluated AppealMod in only one community on Reddit. \add{Furthermore, the AppealMod design is based on interactions with a handful of moderators. While the experience of these moderators is vast and varied (all moderators have years of experience and most have moderated multiple subreddits), we cannot ascertain how the AppealMod design will operate in other communities, and whether its impact will be different for communities with different benefits, rules, and moderators.}\remove{While we are confident that our design will function in other communities, the extent of its impact may be different in communities with different benefits and rules.} For instance, r/pics is the one of the largest communities on Reddit, the most popular place for sharing pictures on Reddit. Given its popularity, users who are banned are likely motivated to get back into the community. At the same time, when users are banned from a public Reddit community, they can still see the community's content, but they cannot post anything. This is unlike \add{platforms like} Facebook where users who are banned from a group can no longer see the group or any of its content.\footnote{\url{https://www.facebook.com/help/211909018842184}} \remove{Therefore, banned users on Facebook may demonstrate an even higher motivation to get back into the community.}

\add{One implication of a complete shutout from the community is that more users would be willing to complete the process compared to Reddit users. Therefore, while Reddit moderators only reviewed 30\% of the appeals submitted under AppealMod, Facebook moderators, for instance, might have to review a higher fraction of appeals submitted under AppealMod. Given the higher decision stake for users on platforms where banning leads to a complete disconnect from the community, another implication is that users might be willing to put more effort in their appeals process. So the AppealMod process for such platforms can be potentially expanded to collect more details from the users that would support the moderators’ review.} Overall, the benefits of the community and the implications of its banning functions will impact how willing users are to bear the additional costs associated with the AppealMod process. Future work should explore how interventions on users' effort impact moderators' workload in different contexts.
% In the absence of more studies that can quantify the impact in different situations, we caution our readers against overgeneralizing the quantified impact of our system on moderator workload. 

Third, we used the Perspective API\footnote{\url{https://perspectiveapi.com/}} to automatically classify user appeals as toxic or not. As we briefly noted in Section~\ref{sec:analysis}, Perspective API occassionaly makes erroneous classifications. These errors will equally impact both the control and treatment groups; our comparison between treatment and control is robust to these errors. However, we caution our readers from generalizing the extent to which AppealMod can reduce the proportion of appeals that are toxic or the overall extent of toxicity in user appeals. While we find that approximately 12\% of all user appeals are toxic, and only 5\% of appeals visible under treatment are toxic, the actual number of toxic appeals may be higher or lower depending on the types of Perspective's errors and their distribution. 

%% file: content/discussion.tex
\section{Discussion}

We worked directly with moderators of a large subreddit to design an appeal process that reduced moderators' workloads, honored users' and moderators' need for direct interaction, and minimized negative impacts to the community. We deployed the AppealMod system for 4 months and found that the process effectively met these goals. Moderators needed to review only 30\% of appeals; users were able to directly discuss their behavior with moderators who made the final decision; users who abandoned the process were either insincere in their appeals or toxic in their comments. In this section, we discuss the impact of AppealMod's effort-based moderation technique on different groups of users. Then, we address the importance and feasibility of meeting both users and moderators' needs when improving moderation systems' contestability, including relevant future work. 

% Finally, we reflect on the importance of combining automation and agency for designing effective moderation systems.

\subsection{Effort Asymmetry and Friction in Content Moderation}

One of the key findings from our design process with moderators was the discovery of \emph{effort asymmetry} between moderators and banned users during the ban appeals process. Moderators spent considerable effort reviewing user appeals of their bans, but users could submit an appeal with minimal effort. To address the asymmetry, AppealMod requires users to provide additional details before their appeals are reviewed by human moderators. Appeals from users who abandon the AppealMod process remain hidden from human moderators. By asking users to put more effort into their appeal, we effectively induced \emph{friction} in appealing, which discouraged many users from completing the process. Our results show that moderators reviewed only 30\% of appeals under the AppealMod process. Overall, our system demonstrates that carefully designed friction can reduce moderators' workload. 

An important concern with inducing friction in a process is the unintended negative effects on participation. For instance, if the increased effort uniformly discouraged users from completing the process, moderators would eventually grant fewer appeals. Fewer successful appeals likely prevent many deserving users from rejoining the community. However, the AppealMod process induced a selection effect among users. We found users who abandoned the process were either insincere in their appeals or toxic in their comments. This selection effect reduced the number of appeals reviewed by moderators by 70\% while moderators still granted roughly the same number of appeals. We hypothesize that the increased effort had a differential impact on users. The AppealMod process likely aided users with a sincere appeal by giving them a space to reflect on their behavior and demonstrate awareness of community norms. Other users, for instance those who were angry at the moderators or did not care about community norms, likely found it more challenging to complete the process and abandoned their effort altogether.
 
The success of inducing friction in a moderation process will largely depend on the type and degree of effort required. As is the case with AppealMod process, the increased effort must be relevant and useful to deserving users and discouraging to problematic users. Furthermore, one must ask for the right amount of effort. The AppealMod process included three-open ended questions and one comment labeling task; the median time for completing the process was under five minutes. Asking users to put in more effort will result in more users abandoning the process and exacerbate any negative outcomes. Asking for too little will invite everyone to complete the process and undermine the effectiveness of this approach. Our design process with moderators was crucial in coming up the right type and amount of effort when designing the system.  

% Collaborating with moderators also enabled us to deploy and evaluate the system in a real setting. 

% In our debriefing session with moderators after the experiment period, we found that moderators liked the system and even requested all user appeals to be addressed via the AppealMod system. Such findings show the importance of directly working with moderators when researching tools for supporting their needs. Finally, it is also important to carefully evaluate such techniques in a real setting to account for any unintended negative consequences in addition to their intended positive effects. 

% \subsection{Contestability of Moderation Decisions}

\subsection{Balancing User and Moderator Needs For Improving the Process of Contesting Moderation Decisions}

In this paper, we presented a new system for contesting content moderation decisions, in particular, the decision of banning users from a community. While we designed our system from the moderators' point of view, some elements of our design support users' contestability needs as well. 
For instance, both users and moderators desire to directly interact with each other and maintain human agency over the final decision. In fact, social media users consider communication, especially direct communication with human moderators, as one of the top three avenues for improving their abilities to contest moderation decisions \cite{vaccaroContestabilityContentModeration2021a}. While AppealMod increases the effort required to directly engage with a human moderator, it does not prohibit it. Our findings show that moderators' response rate was higher on completed appeals under the AppealMod process (80.15\%), compared to their response rate on appeals under control condition (60.04\%). The initial effort users put in to complete the AppealMod process likely prompted human moderators to engage with users more frequently.

The AppealMod process can also guide users in completing their appeals by laying out factors that moderators considered in their decision making. Providing scaffolding can address the cold start problem as many users do not know how to write their appeals \cite{vaccaroEndDayFacebook2020}. Doing so can also make the appeals process more transparent and reduce perceived inconsistencies in moderation decisions \cite{myerswestCensoredSuspendedShadowbanned2018b}. Research on procedural fairness finds that transparency along with the feeling that one’s voice has been heard are important components of users' fairness judgements \cite{lind1988social}. Therefore, completing the AppealMod process can improve a user's perceived fairness and overall satisfaction with the system, especially as it leads to more desirable outcomes for the users. 

Finally, researchers have argued for designing the appeals process to improve users' learning and make them more likely to adhere to the community norms \cite{vaccaroEndDayFacebook2020,myerswestCensoredSuspendedShadowbanned2018b}. AppealMod's current design nudges users to develop a better understanding of the community norms by asking them to differentiate between behaviors permitted and prohibited in the community. \add{Future work could explore extending AppealMod's bot to be an interactive chatbot that can answer users' questions about community norms or how to apply them to their particular situations. Such dialectical exchanges between users and decision systems are identified as an important part of supporting users' contestability needs \cite{sarra2020put}.}

However, these benefits for users come at costs. 
Some users may find it frustrating and cumbersome to interact with an automated bot \cite{cxtodayZendeskResearch}, especially since it is the first step before any interaction with human moderators. Research also argues that adding barriers to a process can be detrimental to users' fairness perceptions \cite{leventhal1976should}. While our work centered on addressing moderators' needs, we encourage future work to further explore users' perceptions about the AppealMod process, how it impacts their behavior after their appeals are decided, and compare it with Reddit's existing ban appeals process.

\section{Conclusion}
AppealMod addressed specific challenges moderators' faced in processing ban appeals. Most importantly, it reduced moderators' workload by creating friction that effectively discouraged insincere appeals. AppealMod also minimized the toxic messages moderators were exposed to and maintained moderators' control over appeal decisions. It achieved these gains for moderators while honoring users' and moderators' preferences for direct human engagement in appeals. The primary difference between the existing appeal process and the process under AppealMod is the level of effort users must put forth to submit an appeal. By introducing productive friction in the process, AppealMod increases the costs of each individual appeal while teaching users about the community's rules and its decision-making process. 

The volume of content moderation that is necessary means that some automation will be required to allow moderators to keep up. However, even moderators make mistakes, and their decisions should be contestable. Our design process and experiment illustrate one strategy for co-creating a system for automating aspects of the moderation process and enabling contestibility of moderator decisions. 

\section{Acknowledgements}
We want to express our deepest gratitude to the moderators of r/pics for their time and cooperation. Without their input, this project would not have been possible. Special thanks to Tawanna Dillahunt and Shagun Jhaver for their generous feedback that helped scope the initial design process. This material is based upon work supported by the National Science Foundation under Grant No. 099726.

\section{Contribution Statements}
The authors confirm contribution to the paper as follows: \\
Shubham: Conceptualization, Methodology, Software, Formal analysis, Investigation, Data Curation, Writing - Original Draft (lead) \\
Jane: Writing - Original Draft (supporting) \\
Paul: Conceptualization, Methodology, Writing - Review and Editing, Supervision (supporting) \\
Libby: Conceptualization, Methodology, Resources, Writing - Original Draft (supporting), Supervision (lead), Funding acquisition 

%% file: content/z_appendix.tex
\newpage
\appendix
\section{Appendix}

\subsection{\add{Recruitment Message}}
\label{app:recruitment}
\textit{
\add{Hi, I'm a researcher at the <masked for anonymity> who is curious about exploring new ways to support moderators, potentially with new tools or other kinds of computational assistance. To make my research impactful, I want to hear from mods about what would be truly useful for them.}}

\textit{
\add{As somebody who no longer lives in <masked for anonymity>, the community has brought a lot of nostalgia, especially the recent set of posts about people sharing their experiences after moving from the US. Now as a researcher, I'd love to talk to you and understand more about moderating this community and any requirements that you may have. If you are interested, I can schedule a quick call to chat more about it. I do value your time (since most moderators are overworked) and would like to offer a gift card worth \$20 as an appreciation for your time and effort. Please let me know if you'd be interested by replying to this message.}}

\textit{
\add{You can find more details about the project here: <masked for anonymity>. 
}}

\vspace{10px}

\subsection{\add{Design Study Materials}}

\subsubsection{\add{Design session protocol}}
\label{app:design-protocol}

\begin{itemize}
    \item \add{From our initial interviews, we found that moderators often receive modmail requests from users appealing against moderation decisions?  How often is that the case?}

    \item \add{How many times is it because the user didn’t understand the rule? How many times do the users just want to argue with you?}

    \item \add{What kind of responses do you want to get from the users?}

    \item \add{How would the user responses help you in responding to their request?}

    \item \add{Do you think the ``good'' and ``bad'' requests would be sufficiently different from each other?}

    \item \add{How often do you think users are likely to complete a process like this?}

    \item \add{Do you see any negative effects on the users while doing this extra work? Follow-up examples: will the users be irritated? do you expect any kind of backlash?}

    \item \add{What if users fake their responses? How would you deal with that?}

    \item \add{Can you share any past data on back-and-forth interactions between users and moderators that can help me design this flow?}
    
\end{itemize}

\newpage
 
\begin{figure}[H]
    \centering
    \includegraphics[width=0.8\textwidth]{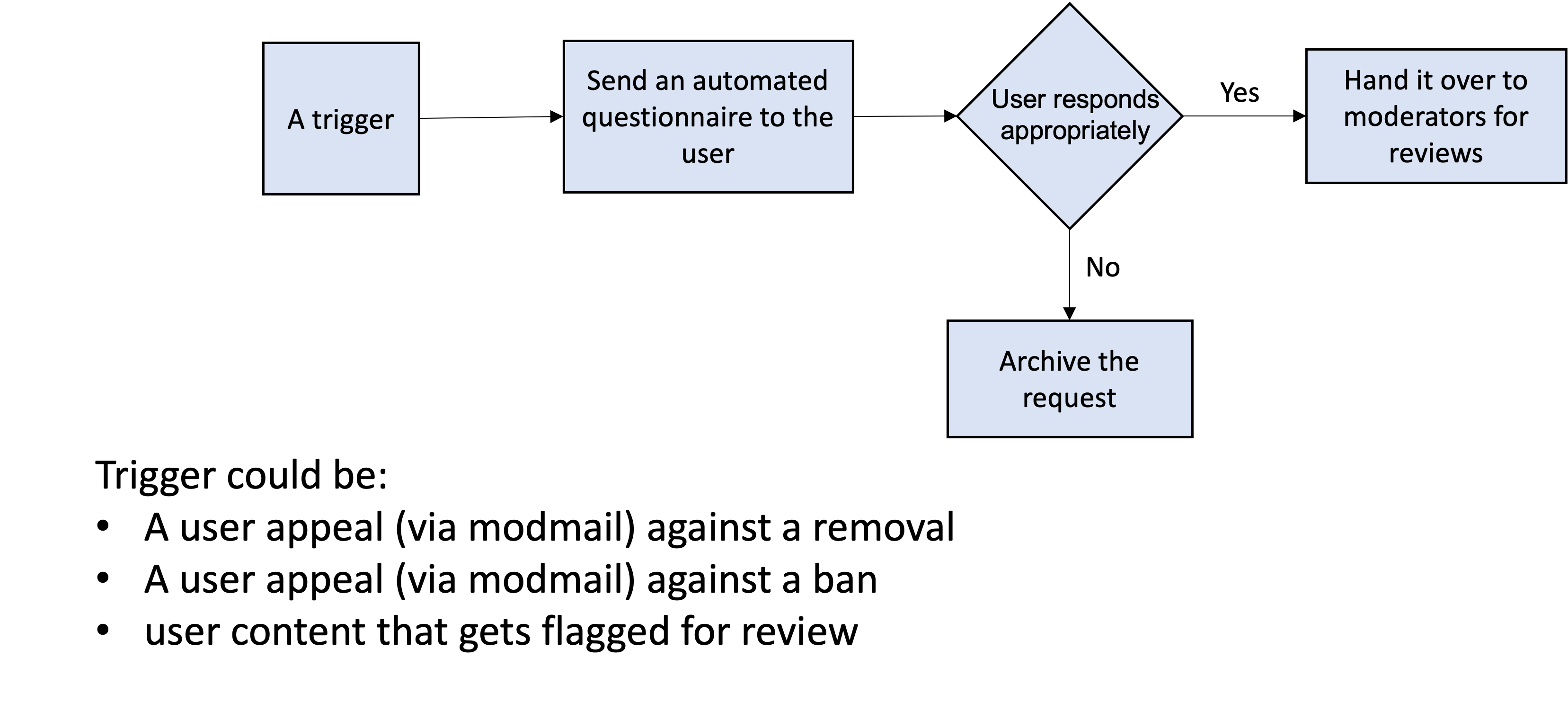}
    \caption{\add{An overview of our proposed design highlighting that users receive automated messages in response to their requests and that some user requests will be hidden from human moderators. The material was sent to the moderators in advance and used during the design sessions.}}
    \label{fig:flowchart}
\end{figure}

% \subsubsection{\add{Questions Typology}}

\begin{figure}[H]
    \centering
    \includegraphics[width=0.8\textwidth]{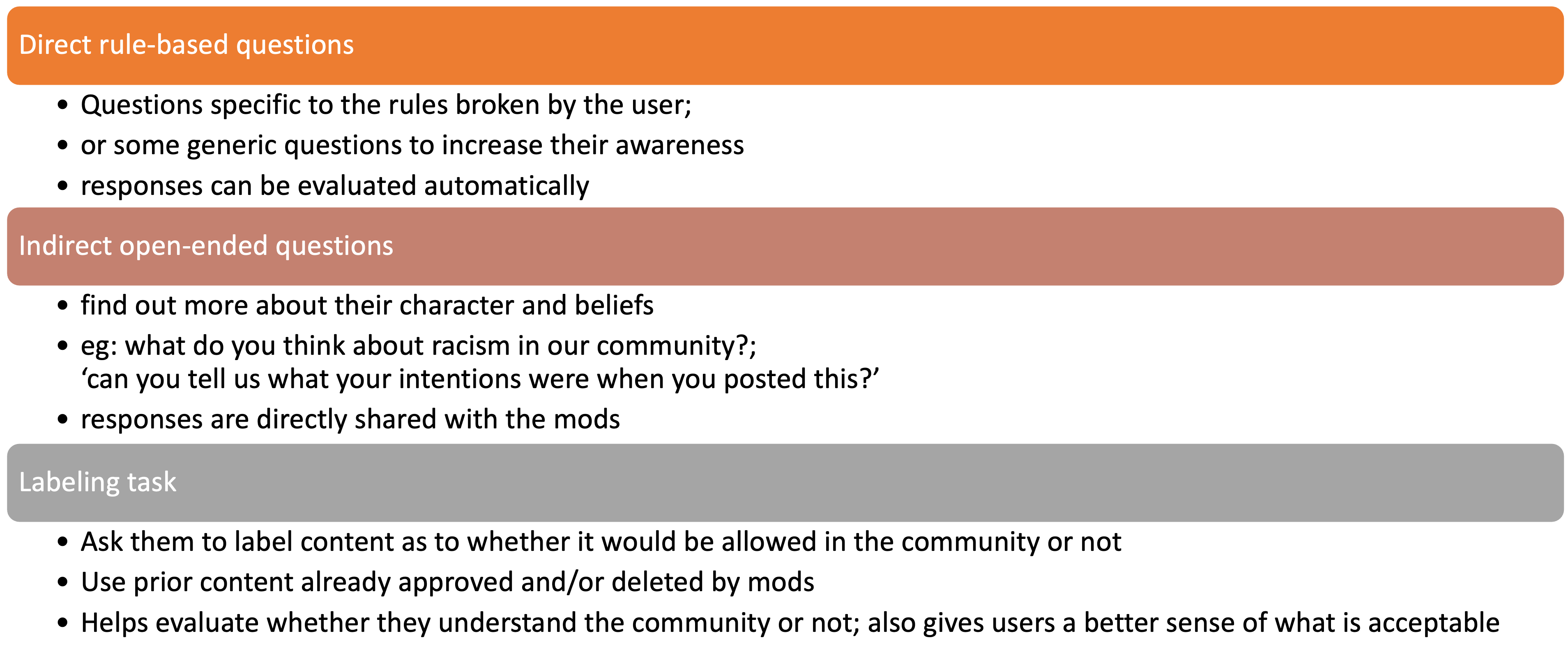}
    \caption{\add{A Typology of different questions that can be asked during the AppealMod process that was sent to the moderators in advance and used during the design sessions. Our final design mostly used open-ended questions along with a comment labeling task.}}
    \label{fig:questions-type}
\end{figure}

% \subsubsection{Mockup conversation}

\begin{figure}[H]
    \centering
    \includegraphics[width=1\textwidth]{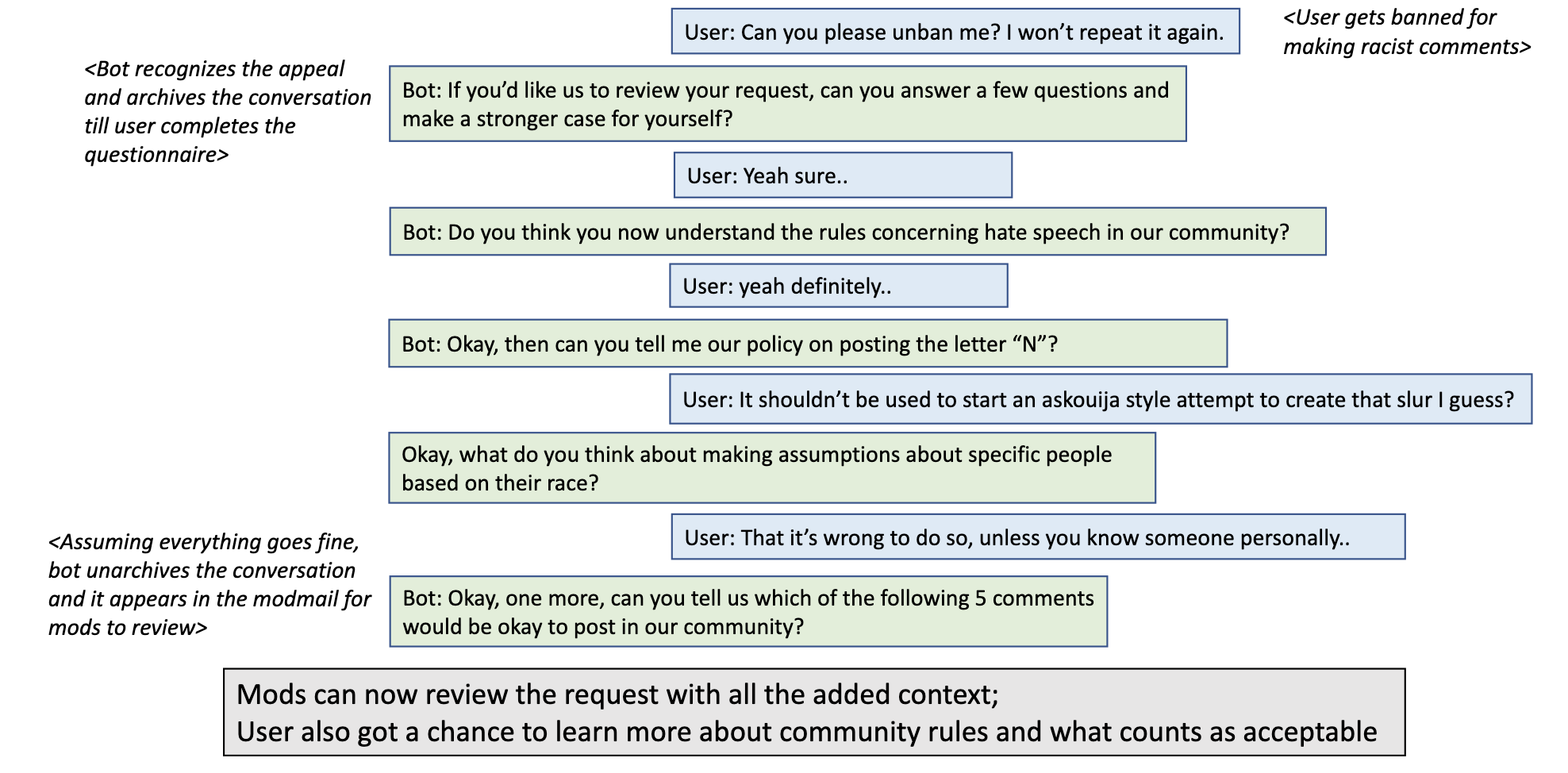}
    \caption{A mockup conversation between our proposed bot and a banned user which was sent to the moderators in advance and used during the design sessions. Following a round of discussions on this prototype, we decided to share the questions with users in a webform instead of directly asking them in the chat. This was done to keep the overall conversation between users and moderators short and organized. }
    \label{fig:prototype}
\end{figure}

\subsection{Questions asked during the AppealMod process}

\input{tables/task-questions.tex}

\newpage
\subsection{Comments used for the comment labeling task during the AppealMod process}

\input{tables/comment-labeling.tex}

\subsection{Impact of different thresholds on toxicity classification}

\input{tables/toxicity-threshold.tex}

\subsection{\add{Protocol for the post-experiment session with moderation team}}
\label{app:post-experiment-protocol} 

\begin{itemize}
    \item \add{Have you observed any recent changes in the ban appeals that you’re receiving? What about appeals you’ve been granting?}

    \item \add{When do you usually review ban appeals? Do you have dedicated time slots, or do you review them as they are submitted?}

    \item \add{Now that you have seen our automated bot in action, what are your initial thoughts about it? Does it help with your review process?}

    \item \add{Select a random set of conversations from control and treatment, and ask the following questions for each of them:}

    \begin{itemize}
        \item \add{As you see this ban appeal, would you respond to it? Why/Why not?}
        
        \item \add{Would you be willing to grant the appeal? Why/Why not?}

        \item \add{If treatment conversation: Is there something specific you’d look for in the form responses?}

        \item \add{If treatment: Did the user’s response help you in any way?}

        \item \add{If treatment: Did it affect your decision? Why/Why not?}

        \item \add{Are there other factors that impacted your decision? Follow up: what about who issued the initial ban? What about the ban reason? }

    \end{itemize}

    \item \add{Finally, any changes you’d like to make in the design?}

\end{itemize}

%% file: tables/task-questions.tex
\begin{table}[h!]
\resizebox{0.9\textwidth}{!}{
\begin{tabular}{l|l|c}
\textbf{Design Feature}               & \textbf{Question}                                                                                                                                       & \textbf{Question Type}   \\ \midrule
Behavior reflection          
&  
\begin{tabular}[l]{@{}l@{}} Can you describe your actions that led to the ban and the circumstances that made \\ you act that way?  \end{tabular}
& Open-ended      \\ \cmidrule{2-3} 
                             & What steps will you take to ensure that you don't repeat such actions in future?                                                               & Open-ended      \\ \midrule
% Awareness of community norms 
\begin{tabular}[l]{@{}l@{}}Awareness of\\community norms\end{tabular}
& \begin{tabular}[l]{@{}l@{}}Can you describe the rule you broke in your own words? \\ You may find more details about the rules here: https://new.reddit.com/r/pics/wiki/index  \end{tabular} 
& Open-ended      \\ \cmidrule{2-3} 
                             & Which of the following comment(s) do you think should be allowed in our community?                                                             & Multiple choice
\end{tabular}
}
\vspace{8px}
\caption{Questions asked during the AppealMod process.}
\label{tab:task-questions}

\end{table}

%% file: tables/comment-labeling.tex
\begin{table}[h!]
\resizebox{0.95\textwidth}{!}{
\begin{tabular}{l|l}
\textbf{Comment} (\textit{Content Warning: Below comments include toxic language})                                                                                                                             & \textbf{Allowed?} \\\midrule
Hahaha, not at all, I love a good discussion, especially with a retard on the internet, but you're too dumb                            & No       \\
You’re still a fat fuck, k..ill yourself                                                                                               & No       \\
I’ll hire my hacker friend to doxx you then send a hitman to give you ebola then shoot you in the head and revive you to give you AIDS & No       \\
Even if your mom is a fucking asshole, I believe you that she is, drugging her will NOT do anyone any good                             & Yes      \\
Fuck him. You dont want or need those assholes in your life. Make a new friend                                                         & Yes     
\end{tabular}
}

\vspace{8px}
\caption{Comments used for the comment labeling task in which users were asked to select comment(s) that they think should be allowed in the community. These comments were reviewed by the moderators and they selected two of them as permissible.}
\label{tab:comment-label}

\end{table}

%% file: tables/toxicity-threshold.tex
\begin{table}[h!]
\small
\resizebox{0.9\textwidth} {!}{
\begin{tabular}{p{0.13\textwidth}|p{0.12\textwidth}|p{0.12\textwidth}|p{0.14\textwidth}|p{0.13\textwidth}|p{0.14\textwidth}}
% & Toxic appeals in control & Toxic appeals in treatment & Toxic appeals that completed AppealMod process & Rate of toxicity in control & Rate of toxicity after AppealMod process \\ \hline
& Toxic appeals in control & Toxic appeals in treatment & Toxic appeals that used AppealMod & Rate of toxicity in control & Rate of toxicity post-AppealMod  \\ \hline
Threshold=0.5 & 79                       & 81                         & 8                                              & 18.03\%                     & 6.1\%                                    \\ \hline
Threshold=0.7 & 58                       & 57                         & 5                                              & 13.24\%                     & 3.81\%                                   \\ \hline
Threshold=0.9 & 32                       & 29                         & 2                                              & 7.3\%                       & 1.52\%                                   \\ 
\end{tabular}
}
\vspace{8px}
\caption{Moderators' exposure to toxicity under control and treatment at different toxicity thresholds}
\label{tab:appendix-toxicity-thresh}
\end{table}

% \begin{table}[]
% \begin{tabular}{|l|ll|ll|ll|}
% \hline
%                                                       & \multicolumn{2}{l|}{Threshold=0.5}       & \multicolumn{2}{l|}{Threshold=0.7}       & \multicolumn{2}{l|}{Threshold=0.9}       \\ \hline
%                                                       & \multicolumn{1}{l|}{Control} & Treatment & \multicolumn{1}{l|}{Control} & Treatment & \multicolumn{1}{l|}{Control} & Treatment \\ \hline
% No. of toxic appeals                                  & \multicolumn{1}{l|}{79}      & 81        & \multicolumn{1}{l|}{58}      & 57        & \multicolumn{1}{l|}{32}      & 29        \\ \hline
% Base rate of toxicity in appeals                      & \multicolumn{1}{l|}{18.03\%} & 18.32\%   & \multicolumn{1}{l|}{13.24\%} & 12.9\%    & \multicolumn{1}{l|}{7.3\%}   & 6.56\%    \\ \hline
% No. of toxic appeals that completed AppealMod process & \multicolumn{1}{l|}{}        & 8         & \multicolumn{1}{l|}{}        & 5         & \multicolumn{1}{l|}{}        & 2         \\ \hline
% Final rate of toxicity in appeals after AppealMod     & \multicolumn{1}{l|}{18.03\%} & 6.1\%     & \multicolumn{1}{l|}{13.24\%} & 3.81\%    & \multicolumn{1}{l|}{7.3\%}   & 1.52\%    \\ \hline
% \end{tabular}
% \end{table}

%% file: main.bbl
%%% -*-BibTeX-*-
%%% Do NOT edit. File created by BibTeX with style
%%% ACM-Reference-Format-Journals [18-Jan-2012].

\begin{thebibliography}{61}

%%% ====================================================================
%%% NOTE TO THE USER: you can override these defaults by providing
%%% customized versions of any of these macros before the \bibliography
%%% command.  Each of them MUST provide its own final punctuation,
%%% except for \shownote{}, \showDOI{}, and \showURL{}.  The latter two
%%% do not use final punctuation, in order to avoid confusing it with
%%% the Web address.
%%%
%%% To suppress output of a particular field, define its macro to expand
%%% to an empty string, or better, \unskip, like this:
%%%
%%% \newcommand{\showDOI}[1]{\unskip}   % LaTeX syntax
%%%
%%% \def \showDOI #1{\unskip}           % plain TeX syntax
%%%
%%% ====================================================================

\ifx \showCODEN    \undefined \def \showCODEN     #1{\unskip}     \fi
\ifx \showDOI      \undefined \def \showDOI       #1{#1}\fi
\ifx \showISBNx    \undefined \def \showISBNx     #1{\unskip}     \fi
\ifx \showISBNxiii \undefined \def \showISBNxiii  #1{\unskip}     \fi
\ifx \showISSN     \undefined \def \showISSN      #1{\unskip}     \fi
\ifx \showLCCN     \undefined \def \showLCCN      #1{\unskip}     \fi
\ifx \shownote     \undefined \def \shownote      #1{#1}          \fi
\ifx \showarticletitle \undefined \def \showarticletitle #1{#1}   \fi
\ifx \showURL      \undefined \def \showURL       {\relax}        \fi
% The following commands are used for tagged output and should be
% invisible to TeX
\providecommand\bibfield[2]{#2}
\providecommand\bibinfo[2]{#2}
\providecommand\natexlab[1]{#1}
\providecommand\showeprint[2][]{arXiv:#2}

\bibitem[Almada(2019)]%
        {almada2019human}
\bibfield{author}{\bibinfo{person}{Marco Almada}.}
  \bibinfo{year}{2019}\natexlab{}.
\newblock \showarticletitle{Human intervention in automated decision-making:
  Toward the construction of contestable systems}. In
  \bibinfo{booktitle}{\emph{Proceedings of the Seventeenth International
  Conference on Artificial Intelligence and Law}}. \bibinfo{pages}{2--11}.
\newblock


\bibitem[Alvarez(2019)]%
        {engadgetInstagramWill}
\bibfield{author}{\bibinfo{person}{Edgar Alvarez}.}
  \bibinfo{year}{2019}\natexlab{}.
\newblock \bibinfo{title}{{I}nstagram will soon let you appeal post takedowns |
  {E}ngadget --- engadget.com}.
\newblock
  \bibinfo{howpublished}{\url{https://www.engadget.com/2019-05-07-instagram-appeals-content-review-taken-down-posts.html}}.
\newblock
\newblock
\shownote{[Accessed 06-Jan-2023]}.


\bibitem[Blackwell et~al\mbox{.}(2017)]%
        {blackwell2017classification}
\bibfield{author}{\bibinfo{person}{Lindsay Blackwell}, \bibinfo{person}{Jill
  Dimond}, \bibinfo{person}{Sarita Schoenebeck}, {and} \bibinfo{person}{Cliff
  Lampe}.} \bibinfo{year}{2017}\natexlab{}.
\newblock \showarticletitle{Classification and Its Consequences for Online
  Harassment: Design Insights from HeartMob}.
\newblock \bibinfo{journal}{\emph{Proc. ACM Hum.-Comput. Interact.}}
  \bibinfo{volume}{1}, \bibinfo{number}{CSCW}, Article \bibinfo{articleno}{24}
  (\bibinfo{date}{dec} \bibinfo{year}{2017}), \bibinfo{numpages}{19}~pages.
\newblock
\urldef\tempurl%
\url{https://doi.org/10.1145/3134659}
\showDOI{\tempurl}


\bibitem[Cai et~al\mbox{.}(2019)]%
        {caiHumanCenteredToolsCoping2019a}
\bibfield{author}{\bibinfo{person}{Carrie~J. Cai}, \bibinfo{person}{Emily
  Reif}, \bibinfo{person}{Narayan Hegde}, \bibinfo{person}{Jason Hipp},
  \bibinfo{person}{Been Kim}, \bibinfo{person}{Daniel Smilkov},
  \bibinfo{person}{Martin Wattenberg}, \bibinfo{person}{Fernanda Viegas},
  \bibinfo{person}{Greg~S. Corrado}, \bibinfo{person}{Martin~C. Stumpe}, {and}
  \bibinfo{person}{Michael Terry}.} \bibinfo{year}{2019}\natexlab{}.
\newblock \showarticletitle{Human-{{Centered Tools}} for {{Coping}} with
  {{Imperfect Algorithms During Medical Decision-Making}}}. In
  \bibinfo{booktitle}{\emph{Proceedings of the 2019 {{CHI Conference}} on
  {{Human Factors}} in {{Computing Systems}}}}. \bibinfo{publisher}{{ACM}},
  \bibinfo{address}{{Glasgow Scotland Uk}}, \bibinfo{pages}{1--14}.
\newblock
\showISBNx{978-1-4503-5970-2}
\urldef\tempurl%
\url{https://doi.org/10.1145/3290605.3300234}
\showDOI{\tempurl}


\bibitem[Cai and Wohn(2019)]%
        {caiCategorizingLiveStreaming2019a}
\bibfield{author}{\bibinfo{person}{Jie Cai} {and}
  \bibinfo{person}{Donghee~Yvette Wohn}.} \bibinfo{year}{2019}\natexlab{}.
\newblock \showarticletitle{Categorizing {{Live Streaming Moderation Tools}}:
  {{An Analysis}} of {{Twitch}}}.
\newblock \bibinfo{journal}{\emph{International Journal of Interactive
  Communication Systems and Technologies}} \bibinfo{volume}{9},
  \bibinfo{number}{2} (\bibinfo{date}{July} \bibinfo{year}{2019}),
  \bibinfo{pages}{36--50}.
\newblock
\showISSN{2155-4218, 2155-4226}
\urldef\tempurl%
\url{https://doi.org/10.4018/IJICST.2019070103}
\showDOI{\tempurl}


\bibitem[Chandrasekharan et~al\mbox{.}(2019)]%
        {chandrasekharanCrossmodCrossCommunityLearningbased2019b}
\bibfield{author}{\bibinfo{person}{Eshwar Chandrasekharan},
  \bibinfo{person}{Chaitrali Gandhi}, \bibinfo{person}{Matthew~Wortley
  Mustelier}, {and} \bibinfo{person}{Eric Gilbert}.}
  \bibinfo{year}{2019}\natexlab{}.
\newblock \showarticletitle{Crossmod: {{A Cross-Community Learning-based
  System}} to {{Assist Reddit Moderators}}}.
\newblock \bibinfo{journal}{\emph{Proceedings of the ACM on Human-Computer
  Interaction}} \bibinfo{volume}{3}, \bibinfo{number}{CSCW}
  (\bibinfo{date}{Nov.} \bibinfo{year}{2019}), \bibinfo{pages}{1--30}.
\newblock
\showISSN{2573-0142}
\urldef\tempurl%
\url{https://doi.org/10.1145/3359276}
\showDOI{\tempurl}


\bibitem[Chandrasekharan et~al\mbox{.}(2022)]%
        {chandrasekharan2022quarantined}
\bibfield{author}{\bibinfo{person}{Eshwar Chandrasekharan},
  \bibinfo{person}{Shagun Jhaver}, \bibinfo{person}{Amy Bruckman}, {and}
  \bibinfo{person}{Eric Gilbert}.} \bibinfo{year}{2022}\natexlab{}.
\newblock \showarticletitle{Quarantined! Examining the effects of a
  community-wide moderation intervention on Reddit}.
\newblock \bibinfo{journal}{\emph{ACM Transactions on Computer-Human
  Interaction (TOCHI)}} \bibinfo{volume}{29}, \bibinfo{number}{4}
  (\bibinfo{year}{2022}), \bibinfo{pages}{1--26}.
\newblock


\bibitem[Cheves(2018)]%
        {outDangerousTrend}
\bibfield{author}{\bibinfo{person}{Alexander Cheves}.}
  \bibinfo{year}{2018}\natexlab{}.
\newblock \bibinfo{title}{{T}he {D}angerous {T}rend of {L}{G}{B}{T}{Q}+
  {C}ensorship on the {I}nternet}.
\newblock
  \bibinfo{howpublished}{\url{https://www.out.com/out-exclusives/2018/12/06/dangerous-trend-lgbtq-censorship-internet}}.
\newblock
\newblock
\shownote{[Accessed 06-Jan-2023]}.


\bibitem[Dosono and Semaan(2019)]%
        {dosonoModerationPracticesEmotional2019}
\bibfield{author}{\bibinfo{person}{Bryan Dosono} {and} \bibinfo{person}{Bryan
  Semaan}.} \bibinfo{year}{2019}\natexlab{}.
\newblock \showarticletitle{Moderation {{Practices}} as {{Emotional Labor}} in
  {{Sustaining Online Communities}}: {{The Case}} of {{AAPI Identity Work}} on
  {{Reddit}}}. In \bibinfo{booktitle}{\emph{Proceedings of the 2019 {{CHI
  Conference}} on {{Human Factors}} in {{Computing Systems}}}}
  \emph{(\bibinfo{series}{{{CHI}} '19})}. \bibinfo{publisher}{{Association for
  Computing Machinery}}, \bibinfo{address}{{New York, NY, USA}},
  \bibinfo{pages}{1--13}.
\newblock
\showISBNx{978-1-4503-5970-2}
\urldef\tempurl%
\url{https://doi.org/10.1145/3290605.3300372}
\showDOI{\tempurl}


\bibitem[Dwork and Naor(1992)]%
        {dwork1992pricing}
\bibfield{author}{\bibinfo{person}{Cynthia Dwork} {and} \bibinfo{person}{Moni
  Naor}.} \bibinfo{year}{1992}\natexlab{}.
\newblock \showarticletitle{Pricing via processing or combatting junk mail}. In
  \bibinfo{booktitle}{\emph{Annual international cryptology conference}}.
  Springer, \bibinfo{pages}{139--147}.
\newblock


\bibitem[Enders et~al\mbox{.}(2008)]%
        {enders2008long}
\bibfield{author}{\bibinfo{person}{Albrecht Enders}, \bibinfo{person}{Harald
  Hungenberg}, \bibinfo{person}{Hans-Peter Denker}, {and}
  \bibinfo{person}{Sebastian Mauch}.} \bibinfo{year}{2008}\natexlab{}.
\newblock \showarticletitle{The long tail of social networking.: Revenue models
  of social networking sites}.
\newblock \bibinfo{journal}{\emph{European Management Journal}}
  \bibinfo{volume}{26}, \bibinfo{number}{3} (\bibinfo{year}{2008}),
  \bibinfo{pages}{199--211}.
\newblock


\bibitem[Fay and Proschan(2010)]%
        {fay2010wilcoxon}
\bibfield{author}{\bibinfo{person}{Michael~P Fay} {and}
  \bibinfo{person}{Michael~A Proschan}.} \bibinfo{year}{2010}\natexlab{}.
\newblock \showarticletitle{Wilcoxon-Mann-Whitney or t-test? On assumptions for
  hypothesis tests and multiple interpretations of decision rules}.
\newblock \bibinfo{journal}{\emph{Statistics surveys}}  \bibinfo{volume}{4}
  (\bibinfo{year}{2010}), \bibinfo{pages}{1}.
\newblock


\bibitem[Gilbert(2020)]%
        {gilbertRunWorldLargest2020a}
\bibfield{author}{\bibinfo{person}{Sarah~A. Gilbert}.}
  \bibinfo{year}{2020}\natexlab{}.
\newblock \showarticletitle{"{{I}} Run the World's Largest Historical Outreach
  Project and It's on a Cesspool of a Website." {{Moderating}} a {{Public
  Scholarship Site}} on {{Reddit}}: {{A Case Study}} of r/{{AskHistorians}}}.
\newblock \bibinfo{journal}{\emph{Proceedings of the ACM on Human-Computer
  Interaction}} \bibinfo{volume}{4}, \bibinfo{number}{CSCW1}
  (\bibinfo{date}{May} \bibinfo{year}{2020}), \bibinfo{pages}{1--27}.
\newblock
\showISSN{2573-0142}
\urldef\tempurl%
\url{https://doi.org/10.1145/3392822}
\showDOI{\tempurl}


\bibitem[Gillespie(2018)]%
        {gillespieCustodiansInternetPlatforms2018}
\bibfield{author}{\bibinfo{person}{Tarleton Gillespie}.}
  \bibinfo{year}{2018}\natexlab{}.
\newblock \bibinfo{booktitle}{\emph{Custodians of the {{Internet}}:
  {{Platforms}}, {{Content Moderation}}, and the {{Hidden Decisions That Shape
  Social Media}}}}.
\newblock \bibinfo{publisher}{{Yale University Press}}, \bibinfo{address}{{New
  Haven, UNITED STATES}}.
\newblock
\showISBNx{978-0-300-23502-9}


\bibitem[Grimmelmann(2017)]%
        {grimmelmannVirtuesModeration2017}
\bibfield{author}{\bibinfo{person}{James Grimmelmann}.}
  \bibinfo{year}{2017}\natexlab{}.
\newblock \bibinfo{booktitle}{\emph{The {{Virtues}} of {{Moderation}}}}.
\newblock \bibinfo{type}{Preprint}. \bibinfo{institution}{{LawArXiv}}.
\newblock
\urldef\tempurl%
\url{https://doi.org/10.31228/osf.io/qwxf5}
\showDOI{\tempurl}


\bibitem[Haimson et~al\mbox{.}(2021)]%
        {haimson2021disproportionate}
\bibfield{author}{\bibinfo{person}{Oliver~L Haimson}, \bibinfo{person}{Daniel
  Delmonaco}, \bibinfo{person}{Peipei Nie}, {and} \bibinfo{person}{Andrea
  Wegner}.} \bibinfo{year}{2021}\natexlab{}.
\newblock \showarticletitle{Disproportionate removals and differing content
  moderation experiences for conservative, transgender, and black social media
  users: Marginalization and moderation gray areas}.
\newblock \bibinfo{journal}{\emph{Proceedings of the ACM on Human-Computer
  Interaction}} \bibinfo{volume}{5}, \bibinfo{number}{CSCW2}
  (\bibinfo{year}{2021}), \bibinfo{pages}{1--35}.
\newblock


\bibitem[Heer(2019)]%
        {heerAgencyAutomationDesigning2019a}
\bibfield{author}{\bibinfo{person}{Jeffrey Heer}.}
  \bibinfo{year}{2019}\natexlab{}.
\newblock \showarticletitle{Agency plus Automation: {{Designing}} Artificial
  Intelligence into Interactive Systems}.
\newblock \bibinfo{journal}{\emph{Proceedings of the National Academy of
  Sciences}} \bibinfo{volume}{116}, \bibinfo{number}{6} (\bibinfo{date}{Feb.}
  \bibinfo{year}{2019}), \bibinfo{pages}{1844--1850}.
\newblock
\urldef\tempurl%
\url{https://doi.org/10.1073/pnas.1807184115}
\showDOI{\tempurl}


\bibitem[Holpuch(2015)]%
        {holpuch2015facebook}
\bibfield{author}{\bibinfo{person}{Amanda Holpuch}.}
  \bibinfo{year}{2015}\natexlab{}.
\newblock \showarticletitle{Facebook still suspending Native Americans over
  “real name” policy}.
\newblock \bibinfo{journal}{\emph{The Guardian}}  \bibinfo{volume}{16}
  (\bibinfo{year}{2015}).
\newblock
\urldef\tempurl%
\url{https://www.theguardian.com/technology/2015/feb/16/facebook-real-name-policy-suspends-native-americans}
\showURL{%
\tempurl}


\bibitem[Holton(2007)]%
        {holton2007coding}
\bibfield{author}{\bibinfo{person}{Judith~A Holton}.}
  \bibinfo{year}{2007}\natexlab{}.
\newblock \showarticletitle{The coding process and its challenges}.
\newblock \bibinfo{journal}{\emph{The Sage handbook of grounded theory}}
  \bibinfo{volume}{3} (\bibinfo{year}{2007}), \bibinfo{pages}{265--289}.
\newblock


\bibitem[Humphreys(2015)]%
        {humphreys2015reflections}
\bibfield{author}{\bibinfo{person}{Macartan Humphreys}.}
  \bibinfo{year}{2015}\natexlab{}.
\newblock \showarticletitle{Reflections on the ethics of social
  experimentation}.
\newblock \bibinfo{journal}{\emph{Journal of Globalization and Development}}
  \bibinfo{volume}{6}, \bibinfo{number}{1} (\bibinfo{year}{2015}),
  \bibinfo{pages}{87--112}.
\newblock
\urldef\tempurl%
\url{https://www.degruyter.com/document/doi/10.1515/jgd-2014-0016/html}
\showURL{%
\tempurl}


\bibitem[Im et~al\mbox{.}(2020)]%
        {im2020synthesized}
\bibfield{author}{\bibinfo{person}{Jane Im}, \bibinfo{person}{Sonali Tandon},
  \bibinfo{person}{Eshwar Chandrasekharan}, \bibinfo{person}{Taylor Denby},
  {and} \bibinfo{person}{Eric Gilbert}.} \bibinfo{year}{2020}\natexlab{}.
\newblock \showarticletitle{Synthesized Social Signals: Computationally-Derived
  Social Signals from Account Histories}. In
  \bibinfo{booktitle}{\emph{Proceedings of the 2020 CHI Conference on Human
  Factors in Computing Systems}} (Honolulu, HI, USA)
  \emph{(\bibinfo{series}{CHI '20})}. \bibinfo{publisher}{Association for
  Computing Machinery}, \bibinfo{address}{New York, NY, USA},
  \bibinfo{pages}{1–12}.
\newblock
\showISBNx{9781450367080}
\urldef\tempurl%
\url{https://doi.org/10.1145/3313831.3376383}
\showDOI{\tempurl}


\bibitem[Inc.(2021)]%
        {redditincTransparencyReport}
\bibfield{author}{\bibinfo{person}{Reddit Inc.}}
  \bibinfo{year}{2021}\natexlab{}.
\newblock \bibinfo{title}{{T}ransparency {R}eport 2021 - {R}eddit}.
\newblock
  \bibinfo{howpublished}{\url{https://www.redditinc.com/policies/transparency-report-2021-2/}}.
\newblock
\newblock
\shownote{[Accessed 14-Jan-2023]}.


\bibitem[Jhaver et~al\mbox{.}(2019)]%
        {jhaverHumanMachineCollaborationContent2019b}
\bibfield{author}{\bibinfo{person}{Shagun Jhaver}, \bibinfo{person}{Iris
  Birman}, \bibinfo{person}{Eric Gilbert}, {and} \bibinfo{person}{Amy
  Bruckman}.} \bibinfo{year}{2019}\natexlab{}.
\newblock \showarticletitle{Human-{{Machine Collaboration}} for {{Content
  Regulation}}: {{The Case}} of {{Reddit Automoderator}}}.
\newblock \bibinfo{journal}{\emph{ACM Transactions on Computer-Human
  Interaction}} \bibinfo{volume}{26}, \bibinfo{number}{5} (\bibinfo{date}{Oct.}
  \bibinfo{year}{2019}), \bibinfo{pages}{1--35}.
\newblock
\showISSN{1073-0516, 1557-7325}
\urldef\tempurl%
\url{https://doi.org/10.1145/3338243}
\showDOI{\tempurl}


\bibitem[Jhaver et~al\mbox{.}(2022)]%
        {jhaverDesigningWordFilter2022}
\bibfield{author}{\bibinfo{person}{Shagun Jhaver}, \bibinfo{person}{Quan~Ze
  Chen}, \bibinfo{person}{Detlef Knauss}, {and} \bibinfo{person}{Amy~X.
  Zhang}.} \bibinfo{year}{2022}\natexlab{}.
\newblock \showarticletitle{Designing {{Word Filter Tools}} for {{Creator-led
  Comment Moderation}}}. In \bibinfo{booktitle}{\emph{{{CHI Conference}} on
  {{Human Factors}} in {{Computing Systems}}}}. \bibinfo{publisher}{{ACM}},
  \bibinfo{address}{{New Orleans LA USA}}, \bibinfo{pages}{1--21}.
\newblock
\showISBNx{978-1-4503-9157-3}
\urldef\tempurl%
\url{https://doi.org/10.1145/3491102.3517505}
\showDOI{\tempurl}


\bibitem[Jiang(2020)]%
        {jiang2020toward}
\bibfield{author}{\bibinfo{person}{Aaron~Jialun Jiang}.}
  \bibinfo{year}{2020}\natexlab{}.
\newblock \showarticletitle{Toward A Multi-stakeholder Perspective For
  Improving Online Content Moderation}.
\newblock  (\bibinfo{year}{2020}).
\newblock


\bibitem[Jiang et~al\mbox{.}(2022a)]%
        {jiang2022trade}
\bibfield{author}{\bibinfo{person}{Jialun~Aaron Jiang}, \bibinfo{person}{Peipei
  Nie}, \bibinfo{person}{Jed~R Brubaker}, {and} \bibinfo{person}{Casey
  Fiesler}.} \bibinfo{year}{2022}\natexlab{a}.
\newblock \showarticletitle{A Trade-off-centered Framework of Content
  Moderation}.
\newblock \bibinfo{journal}{\emph{arXiv preprint arXiv:2206.03450}}
  (\bibinfo{year}{2022}).
\newblock


\bibitem[Jiang et~al\mbox{.}(2022b)]%
        {jiangTradeoffcenteredFrameworkContent2022}
\bibfield{author}{\bibinfo{person}{Jialun~Aaron Jiang}, \bibinfo{person}{Peipei
  Nie}, \bibinfo{person}{Jed~R. Brubaker}, {and} \bibinfo{person}{Casey
  Fiesler}.} \bibinfo{year}{2022}\natexlab{b}.
\newblock \bibinfo{title}{A {{Trade-off-centered Framework}} of {{Content
  Moderation}}}.
\newblock
\newblock
\urldef\tempurl%
\url{https://doi.org/10.1145/3534929}
\showDOI{\tempurl}
\showeprint[arxiv]{2206.03450}~[cs]


\bibitem[Julia~Angwin(2017)]%
        {propublicaFacebooksSecret}
\bibfield{author}{\bibinfo{person}{Hannes~Grassegger Julia~Angwin}.}
  \bibinfo{year}{2017}\natexlab{}.
\newblock \bibinfo{title}{{F}acebook’s {S}ecret {C}ensorship {R}ules
  {P}rotect {W}hite {M}en {F}rom {H}ate {S}peech {B}ut {N}ot {B}lack
  {C}hildren}.
\newblock
  \bibinfo{howpublished}{\url{https://www.propublica.org/article/facebook-hate-speech-censorship-internal-documents-algorithms}}.
\newblock
\newblock
\shownote{[Accessed 06-Jan-2023]}.


\bibitem[Kraut and Resnick(2012)]%
        {krautBuildingSuccessfulOnline2012}
\bibfield{author}{\bibinfo{person}{Robert~E. Kraut} {and} \bibinfo{person}{Paul
  Resnick}.} \bibinfo{year}{2012}\natexlab{}.
\newblock \bibinfo{booktitle}{\emph{Building {{Successful Online Communities}}:
  {{Evidence-Based Social Design}}}}.
\newblock
\urldef\tempurl%
\url{https://doi.org/10.7551/mitpress/8472.001.0001}
\showDOI{\tempurl}


\bibitem[Leventhal(1976)]%
        {leventhal1976should}
\bibfield{author}{\bibinfo{person}{Gerald~S Leventhal}.}
  \bibinfo{year}{1976}\natexlab{}.
\newblock \showarticletitle{What Should be Done Equity Theory? New Approaches
  to the Study of Fairness in Social Relations}.
\newblock \bibinfo{journal}{\emph{Washington DC: National Science Foundation}}
  (\bibinfo{year}{1976}).
\newblock


\bibitem[Levy(2014)]%
        {levy2014facebook}
\bibfield{author}{\bibinfo{person}{Karyne Levy}.}
  \bibinfo{year}{2014}\natexlab{}.
\newblock \showarticletitle{Facebook Apologizes for ‘Real Name’Policy That
  Forced Drag Queens To Change Their Profiles}.
\newblock \bibinfo{journal}{\emph{Business Insider}}  \bibinfo{volume}{1}
  (\bibinfo{year}{2014}).
\newblock
\urldef\tempurl%
\url{https://www.businessinsider.com/facebook-apologizes-for-real-name-policy-2014-10}
\showURL{%
\tempurl}


\bibitem[Li et~al\mbox{.}(2022)]%
        {liAllThatHappening2022}
\bibfield{author}{\bibinfo{person}{Hanlin Li}, \bibinfo{person}{Brent Hecht},
  {and} \bibinfo{person}{Stevie Chancellor}.} \bibinfo{year}{2022}\natexlab{}.
\newblock \showarticletitle{All {{That}}'s {{Happening}} behind the {{Scenes}}:
  {{Putting}} the {{Spotlight}} on {{Volunteer Moderator Labor}} in
  {{Reddit}}}.
\newblock \bibinfo{journal}{\emph{Proceedings of the International AAAI
  Conference on Web and Social Media}}  \bibinfo{volume}{16}
  (\bibinfo{date}{May} \bibinfo{year}{2022}), \bibinfo{pages}{584--595}.
\newblock
\showISSN{2334-0770}
\urldef\tempurl%
\url{https://doi.org/10.1609/icwsm.v16i1.19317}
\showDOI{\tempurl}


\bibitem[Lind and Tyler(1988)]%
        {lind1988social}
\bibfield{author}{\bibinfo{person}{E~Allan Lind} {and} \bibinfo{person}{Tom~R
  Tyler}.} \bibinfo{year}{1988}\natexlab{}.
\newblock \bibinfo{booktitle}{\emph{The social psychology of procedural
  justice}}.
\newblock \bibinfo{publisher}{Springer Science \& Business Media}.
\newblock


\bibitem[Liu et~al\mbox{.}(2018)]%
        {liuForecastingPresenceIntensity2018}
\bibfield{author}{\bibinfo{person}{Ping Liu}, \bibinfo{person}{Joshua
  Guberman}, \bibinfo{person}{Libby Hemphill}, {and} \bibinfo{person}{Aron
  Culotta}.} \bibinfo{year}{2018}\natexlab{}.
\newblock \showarticletitle{Forecasting the Presence and Intensity of Hostility
  on {{Instagram}} Using Linguistic and Social Features}.
\newblock \bibinfo{journal}{\emph{arXiv:1804.06759 [cs]}}
  (\bibinfo{date}{April} \bibinfo{year}{2018}).
\newblock
\showeprint[arxiv]{1804.06759}~[cs]


\bibitem[Lo(2018)]%
        {loWhenAllYou2018}
\bibfield{author}{\bibinfo{person}{Claudia (Claudia Wai~Yu) Lo}.}
  \bibinfo{year}{2018}\natexlab{}.
\newblock \emph{\bibinfo{title}{When All You Have Is a Banhammer : The Social
  and Communicative Work of {{Volunteer}} Moderators}}.
\newblock Thesis. \bibinfo{school}{Massachusetts Institute of Technology}.
\newblock


\bibitem[Lyons et~al\mbox{.}(2021)]%
        {lyons2021conceptualising}
\bibfield{author}{\bibinfo{person}{Henrietta Lyons}, \bibinfo{person}{Eduardo
  Velloso}, {and} \bibinfo{person}{Tim Miller}.}
  \bibinfo{year}{2021}\natexlab{}.
\newblock \showarticletitle{Conceptualising contestability: Perspectives on
  contesting algorithmic decisions}.
\newblock \bibinfo{journal}{\emph{Proceedings of the ACM on Human-Computer
  Interaction}} \bibinfo{volume}{5}, \bibinfo{number}{CSCW1}
  (\bibinfo{year}{2021}), \bibinfo{pages}{1--25}.
\newblock


\bibitem[Mahar et~al\mbox{.}(2018)]%
        {mahar2018squadbox}
\bibfield{author}{\bibinfo{person}{Kaitlin Mahar}, \bibinfo{person}{Amy~X.
  Zhang}, {and} \bibinfo{person}{David Karger}.}
  \bibinfo{year}{2018}\natexlab{}.
\newblock \showarticletitle{Squadbox: A Tool to Combat Email Harassment Using
  Friendsourced Moderation}. In \bibinfo{booktitle}{\emph{Proceedings of the
  2018 CHI Conference on Human Factors in Computing Systems}} (Montreal QC,
  Canada) \emph{(\bibinfo{series}{CHI '18})}. \bibinfo{publisher}{Association
  for Computing Machinery}, \bibinfo{address}{New York, NY, USA},
  \bibinfo{pages}{1–13}.
\newblock
\showISBNx{9781450356206}
\urldef\tempurl%
\url{https://doi.org/10.1145/3173574.3174160}
\showDOI{\tempurl}


\bibitem[Marwick and Miller(2014)]%
        {marwick2014online}
\bibfield{author}{\bibinfo{person}{Alice~E Marwick} {and} \bibinfo{person}{Ross
  Miller}.} \bibinfo{year}{2014}\natexlab{}.
\newblock \showarticletitle{Online harassment, defamation, and hateful speech:
  A primer of the legal landscape}.
\newblock \bibinfo{journal}{\emph{Fordham Center on Law and Information Policy
  Report}} \bibinfo{number}{2} (\bibinfo{year}{2014}).
\newblock
\urldef\tempurl%
\url{https://papers.ssrn.com/sol3/papers.cfm?abstract_id=2447904}
\showURL{%
\tempurl}


\bibitem[Matias(2016)]%
        {matiasGoingDarkSocial2016}
\bibfield{author}{\bibinfo{person}{J.~Nathan Matias}.}
  \bibinfo{year}{2016}\natexlab{}.
\newblock \showarticletitle{Going {{Dark}}: {{Social Factors}} in {{Collective
  Action Against Platform Operators}} in the {{Reddit Blackout}}}. In
  \bibinfo{booktitle}{\emph{Proceedings of the 2016 {{CHI Conference}} on
  {{Human Factors}} in {{Computing Systems}}}} \emph{(\bibinfo{series}{{{CHI}}
  '16})}. \bibinfo{publisher}{{Association for Computing Machinery}},
  \bibinfo{address}{{New York, NY, USA}}, \bibinfo{pages}{1138--1151}.
\newblock
\showISBNx{978-1-4503-3362-7}
\urldef\tempurl%
\url{https://doi.org/10.1145/2858036.2858391}
\showDOI{\tempurl}


\bibitem[Matias(2019)]%
        {matiasCivicLaborVolunteer2019}
\bibfield{author}{\bibinfo{person}{J.~Nathan Matias}.}
  \bibinfo{year}{2019}\natexlab{}.
\newblock \showarticletitle{The {{Civic Labor}} of {{Volunteer Moderators
  Online}}}.
\newblock \bibinfo{journal}{\emph{Social Media + Society}} \bibinfo{volume}{5},
  \bibinfo{number}{2} (\bibinfo{date}{April} \bibinfo{year}{2019}),
  \bibinfo{pages}{205630511983677}.
\newblock
\showISSN{2056-3051, 2056-3051}
\urldef\tempurl%
\url{https://doi.org/10.1177/2056305119836778}
\showDOI{\tempurl}


\bibitem[Merriam et~al\mbox{.}(2002)]%
        {merriam2002introduction}
\bibfield{author}{\bibinfo{person}{Sharan~B Merriam} {et~al\mbox{.}}}
  \bibinfo{year}{2002}\natexlab{}.
\newblock \showarticletitle{Introduction to qualitative research}.
\newblock \bibinfo{journal}{\emph{Qualitative research in practice: Examples
  for discussion and analysis}} \bibinfo{volume}{1}, \bibinfo{number}{1}
  (\bibinfo{year}{2002}), \bibinfo{pages}{1--17}.
\newblock


\bibitem[Mitchell(2022)]%
        {cxtodayZendeskResearch}
\bibfield{author}{\bibinfo{person}{Charlie Mitchell}.}
  \bibinfo{year}{2022}\natexlab{}.
\newblock \bibinfo{title}{{Z}endesk {R}esearch: {C}ustomers {A}re {S}till
  {F}rustrated with {C}hatbots}.
\newblock
  \bibinfo{howpublished}{\url{https://www.cxtoday.com/speech-analytics/customers-frustrated-with-chatbots/}}.
\newblock
\newblock
\shownote{[Accessed 12-Jan-2023]}.


\bibitem[Myers~West(2018)]%
        {myerswestCensoredSuspendedShadowbanned2018b}
\bibfield{author}{\bibinfo{person}{Sarah Myers~West}.}
  \bibinfo{year}{2018}\natexlab{}.
\newblock \showarticletitle{Censored, Suspended, Shadowbanned: {{User}}
  Interpretations of Content Moderation on Social Media Platforms}.
\newblock \bibinfo{journal}{\emph{New Media \& Society}} \bibinfo{volume}{20},
  \bibinfo{number}{11} (\bibinfo{date}{Nov.} \bibinfo{year}{2018}),
  \bibinfo{pages}{4366--4383}.
\newblock
\showISSN{1461-4448}
\urldef\tempurl%
\url{https://doi.org/10.1177/1461444818773059}
\showDOI{\tempurl}


\bibitem[Oestreicher-Singer and Zalmanson(2013)]%
        {oestreicher2013content}
\bibfield{author}{\bibinfo{person}{Gal Oestreicher-Singer} {and}
  \bibinfo{person}{Lior Zalmanson}.} \bibinfo{year}{2013}\natexlab{}.
\newblock \showarticletitle{Content or community? A digital business strategy
  for content providers in the social age}.
\newblock \bibinfo{journal}{\emph{MIS quarterly}} (\bibinfo{year}{2013}),
  \bibinfo{pages}{591--616}.
\newblock


\bibitem[Pan et~al\mbox{.}(2022)]%
        {pan2022comparing}
\bibfield{author}{\bibinfo{person}{Christina~A Pan}, \bibinfo{person}{Sahil
  Yakhmi}, \bibinfo{person}{Tara~P Iyer}, \bibinfo{person}{Evan Strasnick},
  \bibinfo{person}{Amy~X Zhang}, {and} \bibinfo{person}{Michael~S Bernstein}.}
  \bibinfo{year}{2022}\natexlab{}.
\newblock \showarticletitle{Comparing the Perceived Legitimacy of Content
  Moderation Processes: Contractors, Algorithms, Expert Panels, and Digital
  Juries}.
\newblock \bibinfo{journal}{\emph{Proceedings of the ACM on Human-Computer
  Interaction}} \bibinfo{volume}{6}, \bibinfo{number}{CSCW1}
  (\bibinfo{year}{2022}), \bibinfo{pages}{1--31}.
\newblock


\bibitem[Perez(2019)]%
        {techcrunchTwitterLets}
\bibfield{author}{\bibinfo{person}{Sarah Perez}.}
  \bibinfo{year}{2019}\natexlab{}.
\newblock \bibinfo{title}{{T}witter now lets users appeal violations within its
  app}.
\newblock
  \bibinfo{howpublished}{\url{https://techcrunch.com/2019/04/02/twitter-now-lets-users-appeal-violations-within-its-app/}}.
\newblock
\newblock
\shownote{[Accessed 06-Jan-2023]}.


\bibitem[Roberts(2019)]%
        {roberts2019behind}
\bibfield{author}{\bibinfo{person}{Sarah~T Roberts}.}
  \bibinfo{year}{2019}\natexlab{}.
\newblock \bibinfo{booktitle}{\emph{Behind the screen}}.
\newblock \bibinfo{publisher}{Yale University Press}.
\newblock


\bibitem[Sarra(2020)]%
        {sarra2020put}
\bibfield{author}{\bibinfo{person}{Claudio Sarra}.}
  \bibinfo{year}{2020}\natexlab{}.
\newblock \showarticletitle{Put dialectics into the machine: protection against
  automatic-decision-making through a deeper understanding of contestability by
  design}.
\newblock \bibinfo{journal}{\emph{Global Jurist}} \bibinfo{volume}{20},
  \bibinfo{number}{3} (\bibinfo{year}{2020}), \bibinfo{pages}{20200003}.
\newblock


\bibitem[{Sch{\"o}pke-Gonzalez} et~al\mbox{.}(2022)]%
        {schopke-gonzalezWhyVolunteerContent2022}
\bibfield{author}{\bibinfo{person}{Angela~M. {Sch{\"o}pke-Gonzalez}},
  \bibinfo{person}{Shubham Atreja}, \bibinfo{person}{Han~Na Shin},
  \bibinfo{person}{Najmin Ahmed}, {and} \bibinfo{person}{Libby Hemphill}.}
  \bibinfo{year}{2022}\natexlab{}.
\newblock \showarticletitle{Why Do Volunteer Content Moderators Quit?
  {{Burnout}}, Conflict, and Harmful Behaviors}.
\newblock \bibinfo{journal}{\emph{New Media \& Society}} (\bibinfo{date}{Dec.}
  \bibinfo{year}{2022}), \bibinfo{pages}{14614448221138529}.
\newblock
\showISSN{1461-4448}
\urldef\tempurl%
\url{https://doi.org/10.1177/14614448221138529}
\showDOI{\tempurl}


\bibitem[Seering and Kairam(2023)]%
        {seering2023moderates}
\bibfield{author}{\bibinfo{person}{Joseph Seering} {and}
  \bibinfo{person}{Sanjay~R Kairam}.} \bibinfo{year}{2023}\natexlab{}.
\newblock \showarticletitle{Who Moderates on Twitch and What Do They Do?
  Quantifying Practices in Community Moderation on Twitch}.
\newblock \bibinfo{journal}{\emph{Proceedings of the ACM on Human-Computer
  Interaction}} \bibinfo{volume}{7}, \bibinfo{number}{GROUP}
  (\bibinfo{year}{2023}), \bibinfo{pages}{1--18}.
\newblock


\bibitem[Seering et~al\mbox{.}(2022)]%
        {seering2022metaphors}
\bibfield{author}{\bibinfo{person}{Joseph Seering}, \bibinfo{person}{Geoff
  Kaufman}, {and} \bibinfo{person}{Stevie Chancellor}.}
  \bibinfo{year}{2022}\natexlab{}.
\newblock \showarticletitle{Metaphors in moderation}.
\newblock \bibinfo{journal}{\emph{New Media \& Society}} \bibinfo{volume}{24},
  \bibinfo{number}{3} (\bibinfo{year}{2022}), \bibinfo{pages}{621--640}.
\newblock


\bibitem[Seering et~al\mbox{.}(2019)]%
        {seeringModeratorEngagementCommunity2019b}
\bibfield{author}{\bibinfo{person}{Joseph Seering}, \bibinfo{person}{Tony
  Wang}, \bibinfo{person}{Jina Yoon}, {and} \bibinfo{person}{Geoff Kaufman}.}
  \bibinfo{year}{2019}\natexlab{}.
\newblock \showarticletitle{Moderator Engagement and Community Development in
  the Age of Algorithms}.
\newblock \bibinfo{journal}{\emph{New Media \& Society}} \bibinfo{volume}{21},
  \bibinfo{number}{7} (\bibinfo{date}{July} \bibinfo{year}{2019}),
  \bibinfo{pages}{1417--1443}.
\newblock
\showISSN{1461-4448}
\urldef\tempurl%
\url{https://doi.org/10.1177/1461444818821316}
\showDOI{\tempurl}


\bibitem[Steiger et~al\mbox{.}(2021)]%
        {steiger2021psychological}
\bibfield{author}{\bibinfo{person}{Miriah Steiger}, \bibinfo{person}{Timir~J
  Bharucha}, \bibinfo{person}{Sukrit Venkatagiri}, \bibinfo{person}{Martin~J
  Riedl}, {and} \bibinfo{person}{Matthew Lease}.}
  \bibinfo{year}{2021}\natexlab{}.
\newblock \showarticletitle{The psychological well-being of content moderators:
  the emotional labor of commercial moderation and avenues for improving
  support}. In \bibinfo{booktitle}{\emph{Proceedings of the 2021 CHI conference
  on human factors in computing systems}}. \bibinfo{pages}{1--14}.
\newblock


\bibitem[Swatman et~al\mbox{.}(2006)]%
        {swatman2006changing}
\bibfield{author}{\bibinfo{person}{Paula~MC Swatman}, \bibinfo{person}{Cornelia
  Krueger}, {and} \bibinfo{person}{Kornelia Van Der~Beek}.}
  \bibinfo{year}{2006}\natexlab{}.
\newblock \showarticletitle{The changing digital content landscape: An
  evaluation of e-business model development in European online news and
  music}.
\newblock \bibinfo{journal}{\emph{Internet research}} (\bibinfo{year}{2006}).
\newblock


\bibitem[Vaccaro et~al\mbox{.}(2020)]%
        {vaccaroEndDayFacebook2020}
\bibfield{author}{\bibinfo{person}{Kristen Vaccaro}, \bibinfo{person}{Christian
  Sandvig}, {and} \bibinfo{person}{Karrie Karahalios}.}
  \bibinfo{year}{2020}\natexlab{}.
\newblock \showarticletitle{"{{At}} the {{End}} of the {{Day Facebook Does What
  ItWants}}": {{How Users Experience Contesting Algorithmic Content
  Moderation}}}.
\newblock \bibinfo{journal}{\emph{Proceedings of the ACM on Human-Computer
  Interaction}} \bibinfo{volume}{4}, \bibinfo{number}{CSCW2}
  (\bibinfo{date}{Oct.} \bibinfo{year}{2020}), \bibinfo{pages}{167:1--167:22}.
\newblock
\urldef\tempurl%
\url{https://doi.org/10.1145/3415238}
\showDOI{\tempurl}


\bibitem[Vaccaro et~al\mbox{.}(2021)]%
        {vaccaroContestabilityContentModeration2021a}
\bibfield{author}{\bibinfo{person}{Kristen Vaccaro}, \bibinfo{person}{Ziang
  Xiao}, \bibinfo{person}{Kevin Hamilton}, {and} \bibinfo{person}{Karrie
  Karahalios}.} \bibinfo{year}{2021}\natexlab{}.
\newblock \showarticletitle{Contestability {{For Content Moderation}}}.
\newblock \bibinfo{journal}{\emph{Proceedings of the ACM on Human-Computer
  Interaction}} \bibinfo{volume}{5}, \bibinfo{number}{CSCW2}
  (\bibinfo{date}{Oct.} \bibinfo{year}{2021}), \bibinfo{pages}{318:1--318:28}.
\newblock
\urldef\tempurl%
\url{https://doi.org/10.1145/3476059}
\showDOI{\tempurl}


\bibitem[Wohn(2019)]%
        {wohnVolunteerModeratorsTwitch2019}
\bibfield{author}{\bibinfo{person}{Donghee~Yvette Wohn}.}
  \bibinfo{year}{2019}\natexlab{}.
\newblock \showarticletitle{Volunteer {{Moderators}} in {{Twitch Micro
  Communities}}: {{How They Get Involved}}, the {{Roles They Play}}, and the
  {{Emotional Labor They Experience}}}. In
  \bibinfo{booktitle}{\emph{Proceedings of the 2019 {{CHI Conference}} on
  {{Human Factors}} in {{Computing Systems}}}} \emph{(\bibinfo{series}{{{CHI}}
  '19})}. \bibinfo{publisher}{{Association for Computing Machinery}},
  \bibinfo{address}{{New York, NY, USA}}, \bibinfo{pages}{1--13}.
\newblock
\showISBNx{978-1-4503-5970-2}
\urldef\tempurl%
\url{https://doi.org/10.1145/3290605.3300390}
\showDOI{\tempurl}


\bibitem[Wren(2018)]%
        {nprFacebookUpdates}
\bibfield{author}{\bibinfo{person}{Ian Wren}.} \bibinfo{year}{2018}\natexlab{}.
\newblock \bibinfo{title}{{F}acebook {U}pdates {C}ommunity {S}tandards,
  {E}xpands {A}ppeals {P}rocess --- npr.org}.
\newblock
  \bibinfo{howpublished}{\url{https://www.npr.org/2018/04/24/605107093/facebook-updates-community-standards-expands-appeals-process}}.
\newblock
\newblock
\shownote{[Accessed 06-Jan-2023]}.


\bibitem[Wright(2022)]%
        {wright2022automated}
\bibfield{author}{\bibinfo{person}{Lucas Wright}.}
  \bibinfo{year}{2022}\natexlab{}.
\newblock \showarticletitle{Automated Platform Governance Through Visibility
  and Scale: On the Transformational Power of AutoModerator}.
\newblock \bibinfo{journal}{\emph{Social Media+ Society}} \bibinfo{volume}{8},
  \bibinfo{number}{1} (\bibinfo{year}{2022}),
  \bibinfo{pages}{20563051221077020}.
\newblock


\bibitem[Zhang et~al\mbox{.}(2020)]%
        {zhangPolicyKitBuildingGovernance2020b}
\bibfield{author}{\bibinfo{person}{Amy~X. Zhang}, \bibinfo{person}{Grant Hugh},
  {and} \bibinfo{person}{Michael~S. Bernstein}.}
  \bibinfo{year}{2020}\natexlab{}.
\newblock \showarticletitle{{{PolicyKit}}: {{Building Governance}} in {{Online
  Communities}}}. In \bibinfo{booktitle}{\emph{Proceedings of the 33rd {{Annual
  ACM Symposium}} on {{User Interface Software}} and {{Technology}}}}.
  \bibinfo{publisher}{{ACM}}, \bibinfo{address}{{Virtual Event USA}},
  \bibinfo{pages}{365--378}.
\newblock
\showISBNx{978-1-4503-7514-6}
\urldef\tempurl%
\url{https://doi.org/10.1145/3379337.3415858}
\showDOI{\tempurl}


\bibitem[Zong and Matias(2022)]%
        {zong2022bartleby}
\bibfield{author}{\bibinfo{person}{Jonathan Zong} {and}
  \bibinfo{person}{J~Nathan Matias}.} \bibinfo{year}{2022}\natexlab{}.
\newblock \showarticletitle{Bartleby: Procedural and Substantive Ethics in the
  Design of Research Ethics Systems}.
\newblock \bibinfo{journal}{\emph{Social Media+ Society}} \bibinfo{volume}{8},
  \bibinfo{number}{1} (\bibinfo{year}{2022}),
  \bibinfo{pages}{20563051221077021}.
\newblock
\urldef\tempurl%
\url{https://journals.sagepub.com/doi/pdf/10.1177/20563051221077021}
\showURL{%
\tempurl}


\end{thebibliography}
